\documentclass[onecolumn]{IEEEtran}
\usepackage{mathpazo}
\usepackage{times}

\usepackage{amsmath}
\usepackage{amsfonts}
\usepackage{latexsym}
\usepackage{amssymb}
\usepackage{mathabx}
\usepackage{float}

\usepackage{upref}
\usepackage{theorem}
\usepackage{graphicx}
\usepackage{psfrag}
\usepackage{cite}
\usepackage{comment}
\usepackage{color}

\usepackage{algorithm,algpseudocode}

\makeatletter
\newcommand{\removelatexerror}{\let\@latex@error\@gobble}
\makeatletter
\newcommand{\proofpart}[2]{%
	\par
	\addvspace{\medskipamount}%
	\noindent\emph{Part #1: #2}\par\nobreak
	\addvspace{\smallskipamount}%
	\@afterheading
}
\makeatother

\theoremstyle{plain}
\theorembodyfont{\normalfont\slshape}

\newtheorem{thm}{Theorem$\!$}
\newenvironment{theorem}
{\begin{thm}\hspace*{-1ex}{\bf.}}{\end{thm}}

\newtheorem{clm}[thm]{Claim$\!$}

\newtheorem{lem}[thm]{Lemma$\!$}
\newenvironment{lemma}{\begin{lem}\hspace*{-1ex}{\bf.}}{\end{lem}}

\newtheorem{prop}[thm]{Proposition$\!$}

\newtheorem{cor}[thm]{Corollary$\!$}

\newtheorem{defn}[thm]{Definition$\!$}
\newenvironment{definition}{\begin{defn}\hspace*{-1ex}{\bf.}}{\end{defn}}

\newtheorem{xmpl}[thm]{Example$\!$}
\newenvironment{example}{\begin{xmpl}\hspace*{-1ex}{\bf.}}{\hfill $\Box$ \end{xmpl}}

\newtheorem{cnstr}{Construction$\!$}

\newtheorem{rmk}[thm]{Remark$\!$}
\newenvironment{remark}{\begin{rmk}\hspace*{-1ex}{\bf.}}{\end{rmk}}

\newcounter{enumrom}
\renewcommand{\theenumrom}{(\roman{enumrom})}

\makeatletter
\renewcommand{\@endtheorem}{\endtrivlist}
\makeatother

\makeatletter
\renewcommand{\thefigure}{{\@arabic\c@figure}}
\renewcommand{\fnum@figure}{{\bf Figure\,\thefigure}}
\makeatother

\newcommand{\cB}{\mathcal{B}}

\newcommand{\cE}{\mathcal{E}}

\newcommand{\cI}{\mathcal{I}}

\newcommand{\cL}{\mathcal{L}}

\newcommand{\cO}{\mathcal{O}}

\newcommand{\cW}{\mathcal{W}}
\newcommand{\cX}{\mathcal{X}}

\newcommand{\be}{\mathbf{e}}

\newcommand{\bfb}{{\boldsymbol b}}

\newcommand{\mathset}[1]{\left\{#1\right\}}
\newcommand{\abs}[1]{\left|#1\right|}

\newcommand{\floorenv}[1]{\left\lfloor #1 \right\rfloor}
\newcommand{\parenv}[1]{\left( #1 \right)}
\newcommand{\sparenv}[1]{\left[ #1 \right]}

\renewcommand{\leq}{\leqslant}

\renewcommand{\geq}{\geqslant}

\renewcommand{\Bbb}{\mathbb}

\newcommand{\Cref}[1]{Co\-ro\-lla\-ry\,\ref{#1}}

\renewcommand{\Bbb}{\mathbb}

\newcommand{\N}{{\Bbb N}}

\newcommand{\Z}{{\Bbb Z}}
\newcommand{\E}{{\Bbb E}}

\newcommand{\elim}[2]{#2_{\mathrm{el}}(#1)}
\newcommand{\ccap}{\mathsf{cap}}
\newcommand{\fr}{\mathrm{fr}}

\newcommand{\limup}[1]{\lim_{#1\rightarrow\infty}}

\newcommand{\limsupup}[1]{\limsup_{#1\rightarrow\infty}}

\newcommand{\eqdef}{\triangleq}

\newcommand{\1}{\mathbf{1}}

\newcommand{\given}{~\middle|~}

\DeclareMathOperator{\Supp}{Supp}

\DeclareMathOperator{\Cr}{Cr}

\newcommand{\rfrac}[2]{{}^{#1}\!/_{#2}}

\outer\def\proclaim #1. #2\par{\medbreak
 \noindent{\bf#1.\enspace}{\sl#2\par}%
 \ifdim\lastskip<\medskipamount \removelastskip\penalty55\medskip\fi}

\begin{document}

\title{\textbf{Repeat-Free Codes}}

\author{\large Ohad~Elishco,~\IEEEmembership{Member,~IEEE}, Ryan~Gabrys,~\IEEEmembership{Member,~IEEE}, Eitan~Yaakobi,~\IEEEmembership{Senior Member,~IEEE}, and~Muriel~M\'{e}dard,~\IEEEmembership{Fellow,~IEEE}
	\thanks{O. Elishco is with Ben-Gurion University of the Negev, Beer-Sheva, 8410501, Israel (e-mail: \texttt{ohadeli@bgu.ac.il}).}
	\thanks{ M. M\'{e}dard is with Massachusetts Institute of Technology, Cambridge, MA, 02139 (e-mail: \texttt{medard@mit.edu}).}
	\thanks{R. Gabrys is with Spawar Systems Center, San Diego, San Diego, CA, 92115 (e-mail: \texttt{ryan.gabrys@navy.mil}).}
	\thanks{E. Yaakobi is with the Department of Computer Science, Technion --- Israel Institute of Technology, Haifa 32000, Israel (e-mail: \texttt{yaakobi@cs.technion.ac.il}).}
	\thanks{ This paper was presented in part at the IEEE International Symposium on Information Theory (ISIT 2019), Paris, France.}}

\maketitle

\begin{abstract}
In this paper we consider the problem of encoding data into \textit{repeat-free} sequences in which sequences are imposed to contain any $k$-tuple at most once (for predefined $k$). First, the capacity of the repeat-free constraint are calculated. Then, an efficient algorithm, which uses two bits of redundancy, is presented to encode length-$n$ sequences for $k=2+2\log (n)$. This algorithm is then improved to support any value of $k$ of the form $k=a\log (n)$, for $1<a$, while its redundancy is $o(n)$. We also calculate the capacity of repeat-free sequences when combined with local constraints which are given by a constrained system, and the capacity of multi-dimensional repeat-free codes.
\end{abstract}


\section{Introduction}
\label{sec:intro}
%%%%%%%%%%%%%%%%%%%%%%%%%%%%%%%%%%%%%
\IEEEPARstart{R}{}epeat-free sequences represent a generalization of the well-known De-Bruijn sequences in which every length-$k$ substring appears exactly once. De-Bruijn sequences have found applications in areas as diverse as cryptography, pseudo-randomness, and information hiding in wireless communications \cite{Fre1982}. However, one potential drawback to adopting De-Bruijn sequences for representing information is that De-Bruijn sequences have rate at most $1/2$. In this work, we show that by relaxing the condition in which every $k$-tuple appears \emph{exactly} once to appear \emph{at most} once, we can generate codes of asymptotic rate $1$ with efficient encoders and decoders for a variety of parameters.

One motivating application for this work is DNA storage, and, in particular, the reading process of a DNA string. The reading process of a DNA string is as follows. At first, the long string is fragmented into substrings of a shorter length which may be read properly. Then, a multiset of all the short strings is obtained in a form of their frequency. Then, the long DNA string should be reconstructed using only the knowledge of the shorter length substrings. 

There are two common lines of work on DNA storage systems. The first assumes that the data is stored in a living organism. In this case, the major concern is to correct errors which are made by naturally occurring mutations. For analysis of the capacity of mutation strings, see~\cite{FarSchBru15,FarSchBru16,EliFarSchBru16,EliFarSchBru18,LouFarSchBru18} and~\cite{CheChrKiaNgu17,JaiFarSchBru17,DolAna10,LenWacYaa17,Wac2018} for coding and algorithms related works. The second line of work focuses on data storage outside a living organism and is called coding for string reconstruction. The goal of coding for the string reconstruction problem is to encode arbitrary strings into ones that are uniquely reconstructible. This problem is motivated by the reading process of DNA-based data storage, where the stored strings are to be reconstructed from information about substrings appearing in the stored string. This problem motivated a series of papers regarding decoding of sequences from partial information on their substrings \cite{MarSki95,KiaPueMil2016,BenMeySchSmiSto91,Sco97,BatKanKhaMcG2004,DudSch2003,
AchDasMilOrlPan10,AchDasMilOrlPan15,GabMil18}.

In order to ensure unique reconstruction, studies were made on reconstruction of encoded sequences \cite{KiaPueMil2016,ChaChrEzeKia17,Lev01}. One method that guarantees a unique reconstruction is to encode the information sequence to a codeword that does not contain any $k$-tuple more than once. For two positive integers $k<n $, we say that a length-$n$ word $w$ is a \emph{$k$-repeat free word} if every substring of $w$ of length $k$ appears at most once. It is already known that $k$-repeat free words are uniquely reconstructible from their length-$r$ substrings multiset if $r\geq k+1$ \cite{Ukk1992}. Furthermore, an encoding scheme that exploits this property has been recently proposed in~\cite{GabMil18}; however, the encoded words are not strictly repeat free. Thus, studying the \emph{repeat-free constraint} and designing respective efficient encoding and decoding schemes is still an open research problem, which is the primary focus of this paper.

Another important characteristic of the $k$-repeat free sequences is the growth rate of the number of sequences as a function of the length of the sequence. Arguably, one of the most well known families of $k$-repeat free sequences are De-Bruijn sequences of span $k$ which play an important role in this paper. A De-Bruijn sequence of span $k$ is a sequence over a finite alphabet, in which every $k$-tuple appears exactly once. It is clear that every De-Bruijn sequence of span $k$ over an alphabet of size $q$ (which implies that the sequence is of length $q^k+k-1$) is $k$-repeat free \cite{DeB1946}. For the case of De-Bruijn sequences, a closed formula for the number of De-Bruijn sequences of length $q^k+k-1$ exists \cite{DeB1946,Fre1982}. Unfortunately, there is no such formula for the general set of $k$-repeat free sequences.  It is clear that a sequence of length $n$ over an alphabet of size $q$ cannot be $k$-repeat free if $k<\log_q n$. However, the size of $k$-repeat free sequences with $k=a\log_q n$ with $a>1$ has not been fully determined. 

Using union bound arguments it is straightforward to show that the growth rate of the number of $k$-repeat free sequences is $q^n$ when $k = \lceil a \log_q n \rceil$ and $a\geq 2$. On the other hand, from the known enumeration results of De-Bruijn sequences \cite{DeB1946,Fre1982} it follows that the growth rate of $k$-repeat free sequences in the binary case is at least $2^{n/2}$ for $a = 1$ (notice that for $a=1$, the set of length $q^k+k-1$, $k$-repeat free sequences is equal to the set of De-Bruijn sequences of span $k$). It is left to find the growth rate for $1\leq  a < 2$. By applying Lov\'{a}sz local lemma, we show in this paper that the growth rate is roughly $2^n$ for all $a>1$. 

Motivated by several previous works \cite{GabMil18,LevYaa17,LevYaa18}, we address the problem of calculating the capacity of $k$-repeat free sequences of length $n$ where $k=a\log (n)$ with $a>1$. We provide an efficient encoding algorithm that encodes into $k$-repeat free binary words for $k =  2\lceil \log (n)\rceil + 2$ with only two redundancy bits. We also extend this algorithm to the setup where $\log (n) < k < 2 \log (n)$ with asymptotically rate-one algorithm. Both algorithms operate in two phases; in the first phase the information sequence is compressed into some shorter sequence that satisfies the constraint and afterwards this compressed sequence is expanded to ensure that the final output is of length $n$ and yet satisfies the constraint. We also study the capacity of $k$-repeat free sequences which satisfy local constraints such as the combination of the $k$-repeat free constraint and the no-adjacent-zeros constraint (i.e., the $(0,1)$-run-length-limited constraint). We show that the $k$-repeat free constraint does not impose a rate penalty when $a>2\log_{\lambda} 2$ and $\lambda$ is the Perron eigenvalue of the matrix that represents the local constraints.

The capacity results are also generalized to the multidimensional case. While the number of binary De-Bruijn sequences of span $k$ is known, in the multidimensional case, the situation is much more complicated. The analog definition of a De-Bruijn sequence to a multidimensional scenario is called a \emph{De-Bruijn torus}. Not only that the number of De-Bruijn tori is not known, it is not known for which sizes there exists a De-Bruijn torus \cite{HurIsa93,HurIsa95,Ma84}.

The rest of the paper is organized as follows. In Section~\ref{sec:pre}, we present the notation and definitions which are used throughout the paper together with the definition of $k$-repeat free sequences. In Section~\ref{sec:capwdbs}, we present our first result which asserts that the capacity of $k$-repeat free sequences for $k = a \log (n)$ is $1$ whenever $a > 1$. In Section~\ref{sec:alg1}, we present an encoding algorithm for binary sequences of length $n$ with $k=2\lceil \log (n)\rceil+2$ and two bits of redundancy. Next, an encoding algorithm for $k=a\log (n)$ with $1<a\leq 2$ is presented in Section~\ref{sec:alg2}. In Section~\ref{sec:comb}, we calculate the capacity of $k$-repeat free sequences which also satisfy local constraints. In Section~\ref{sec:mdcap}, we generalize the capacity result for $d$-dimensional $k$-repeat free arrays. We conclude in Section~\ref{sec:conc}.

\section{Preliminaries}
\label{sec:pre}
%%%%%%%%%%%%%%%%%%%%%%%%%%%%%%%%%%%%%
Let $\N$ denote the set of natural numbers. For $n\in\N$, we denote by $[n]$ the set $[n]=\mathset{0,1,\dots, n-1}$ and by $[-n]$ the set $[-n]=\mathset{-1,-2,\dots,-n}$. %For any real number $r\in\R$, we denote by $\floorenv{r}$, $\ceilenv{r}$ the floor and ceiling operators, respectively. 
For a set $A$ we use $|A|$ to denote the size of $A$. If $A$ is a subset of a group with a group operation $\bullet$, and if $b$ is any group member, we define $b\bullet A\triangleq \mathset{b\bullet a ~:~ a\in A}$.
\begin{example}
	Let $A=\mathset{1,3,5,6} \subseteq \Z$ and let $b=-1$. Then, 
	\[A+b=\mathset{0,2,4,5},\qquad A\cdot b=\mathset{-1,-3,-5,-6}.\]
\end{example}

Throughout the paper, we use $\Sigma$ to denote a finite alphabet. A word of length $n$ over $\Sigma$, $w=(w_0,\dots,w_{n-1})$ is a sequence of $n$ symbols from $\Sigma$ and is defined as a function from $[n]$ to $\Sigma$. We will sometimes denote words as a sequence of letters which are not separated by commas, i.e., $w=(w_0w_1\dots w_{n-1})$.  We denote by $\Sigma^n$ the set of all functions from $[n]$ to $\Sigma$ and by $\Sigma^*=\bigcup_{n\in\N}\Sigma^n$. For a word $w\in\Sigma^*$, $|w|$ denotes the length of $w$ (i.e., the domain of the function $w$) and $w_i=w(i)$ is simply the $i$th symbol in $w$. 

\begin{definition}
	Let $A,B$ be two sets and let $f:A\to B$ be any function. For a subset $A'\subseteq A$, we denote by $f_{A'}:A'\to B$ the restriction of $f$ to $A'$.
\end{definition}

Since we consider words as functions, for a word $w\in\Sigma^n$ and for a set $A\subseteq [n]$, $w_A$ denotes the restriction of $w$ to the set $A$. In other words, $w_A$ is a word created by taking the symbols from $w$ that appear in the positions in $A$. We say that $u$ is a substring of $w$ if there exists $i\in\N$ such that $w_{i+[|u|]}=(w_i,\dots,w_{i+|u|-1})=u$. If $w,u\in\Sigma^*$ we denote by $wu\in\Sigma^{|w|+|u|}$ the concatenation of $w$ and $u$. We will also use the symbol $w\circ u$ when we would like to emphasize the distinct parts. For $w\in\Sigma^*$ we write $w^{\ell}$ for the concatenation of $w$ with itself $\ell\in\N$ times.  Unless otherwise is mentioned, coordinates of a word $w\in\Sigma^*$ are considered modulo $|w|$. Thus, if $w,u\in\Sigma^*$, we have $wu_{[|w|]}=w$ and $wu_{[-|u|]}=u$.

The main object studied in this paper is a set of words which we call a \emph{system}. Specifically, we focus on systems which are defined using global constraints. One of the main characterizations of a system is given by the number of feasible words of length $n$. To be more specific, we would like to estimate the rate at which the number of length-$n$ words grows with $n$. This value is called the \emph{capacity} of the system and is defined as follows.

\begin{definition}
	Let $\cL\subseteq\Sigma^{*}$ be a system. The \textbf{capacity} of $\cL$ is denoted by $\ccap(\cL)$ and is defined as
	\[ \ccap(\cL)\eqdef \limsupup{n}\frac{1}{n}\log_{|\Sigma|} |\cL\cap\Sigma^{n}|.\]
%	where the logarithm is of base $|\Sigma|$ unless mentioned otherwise.
\end{definition}
In case $q=2$ we will sometime simply write $\log$ instead of $\log_2$.

The systems we consider will be defined mostly using constraints on the number of substring appearances. To this end, we define the notion of empirical frequency. 

\begin{definition}
	Let $w\in\Sigma^{n}$ and $k\leq n$. The \textbf{empirical frequency} of $k$-tuples in $w$ is denoted by $\fr^k_w$ and is defined as follows. For a $k$-tuple, $u\in\Sigma^k$, 
	\[\fr_w^k(u)\eqdef \frac{1}{(n-k+1)}\sum_{m\in [n-k+1]} \mathbb{1}_{u}\parenv{w_{m+[k]}},\]
	where $\mathbb{1}$ denotes the indicator function defined by $\mathbb{1}_a(b)=1$ if $b=a$ and $0$ otherwise. 
	We will sometimes consider $\fr^k_w$ as a vector of length $|\Sigma|^k$ or as a probability distribution.
\end{definition}
The \emph{support} of $\fr^k_w$, denoted by $\Supp (\fr^k_w)$, is the set of all $k$-tuples which appear in $w$.

\begin{example}
	Let $\Sigma$ be the binary alphabet and let $w=(11001010), v=(00111010)\in\Sigma^{8}$. For $k=2$, the empirical frequency of the pairs in $w, v$ is given by $\fr^2_w, \fr^2_v$, respectively. We have that 
	\begin{align*}
	\fr^2_w(01)=\frac{2}{7}, \; \;\fr^2_w(10)=\frac{3}{7}, \; \; \fr^2_w(11)=\fr^2_w(00)=\frac{1}{7}
	\end{align*}
	and 
	\[\fr^2_v(01)=\fr^2_v(10)=\fr^2_{v}(11)=\frac{2}{7},\; \; \fr^2_{v}(00)=\frac{1}{7}.\]
	Both $w,v$ have full support, i.e., 
	\[\Supp(\fr^2_w)=\Supp(\fr^2_v)=\Sigma^2,\] 
	but $\Supp(\fr^5_w)=\mathset{11001,10010,00101,01010}$ and $\Supp(\fr^5_v)=\mathset{00111,01110,11101,11010}$.
\end{example}

One of the most important sets of words related to this work is the set of (one-dimensional) De-Bruijn sequences~\cite{DeB1946}. We follow the non-cyclic definition of De-Bruijn sequences and for a finite alphabet $\Sigma$, and for $1\leq k\in\N$, we say that a word $w$ is a \emph{De-Bruijn word of span $k$} if every $k$-tuple appears in $w$ exactly once. Note that $w$ must be of length $\abs{\Sigma}^k+k-1$ (where the $(k-1)$-suffix equals to the $(k-1)$-prefix), since there are exactly $|\Sigma|^k$ different $k$-tuples. Using our notation, we define the following system.

\begin{definition}
	A word $w\in\Sigma^*$ is called a \textbf{De-Bruijn sequence of span $k$} if every $k$-tuple appears exactly once, i.e., for every $u\in\Sigma^k$, 
	\[\fr^k_w(u)=\frac{1}{|w|-k+1}.\]
	The \textbf{De-Bruijn system} over the alphabet $\Sigma$ with $|\Sigma|=q$ is denoted by $\cB_q$ and is defined as the set of all De-Bruijn sequences (over $\Sigma$) of span $k$ for some $1\leq k\in\N$. In a notational form, a De-Bruijn system over $\Sigma$ is the set 
	\begin{align*}
	&\cB_{q}= \\
	&\mathset{w\in \Sigma^* ~:~ \exists k\in\N\; s.t.\; \forall u\in\Sigma^k,\;\; \fr^k_w(u)=\frac{1}{|w|-k+1}}.
	\end{align*}
\end{definition}
Note that by definition, a De-Bruijn system contains all the De-Bruijn sequences of span $k$, for some $k\in\N$. In fact, a De-Bruijn system contains words of lengths $|\Sigma|^k+k-1$ for some $k$. 

The number of binary De-Bruijn sequences of span $k$ is known due to De-Bruijn himself who used the doubling process to calculate the exact number \cite{DeB1946}. Later, his result was generalized to any alphabet \cite{Fre1982}. For a finite alphabet $\Sigma$ with $|\Sigma|=q$, the number of De-Bruijn sequences of span $k$ is given by
\[\parenv{(q-1)!}^{q^{k-1}}\cdot q^{q^{k-1}-k}.\]
Using this formula, the capacity of the De-Bruijn system can be calculated as follows.
For all $n\neq q^{k-1}+k-1$ for some $k\in\N$ we obtain that $\cB_{q}\cap\Sigma^n=\emptyset$ which implies that $\log_q |\cB_q\cap\Sigma^n|=-\infty$. For $n=q^k+k-1$ for some $k\in\N$ we obtain 
\begin{align*}
\ccap(\cB_q)&= \limsupup{n} \frac{1}{n}\log_q |\cB_q \cap \Sigma^n|\\
&\geq \limup{k} \frac{1}{q^k+k-1}\log_q \parenv{\parenv{(q-1)!}^{q^{k-1}}\cdot q^{q^{k-1}-k}} \\
&= \frac{1}{q}(\log_q \parenv{q!}).
\end{align*}
Hence, $\ccap (\cB_q)=\frac{1}{q}(\log_q \parenv{q!})$. Note that when $q=2$, $\ccap(\cB_{2})=1/2$ and using Stirling's approximation we also obtain that $\limup{q} \ccap(\cB_q)=1$.

\section{Capacity of $k$-Repeat Free Systems}
\label{sec:capwdbs}
In this section we introduce the first system we will consider in this work and calculate the capacity of the system. One may regard this system as a generalization of De-Bruijn systems. Throughout this section and unless stated otherwise, we let $\Sigma$ be a fixed alphabet of size $q$. 
\begin{definition}
	A sequence $w\in\Sigma^{n}$ is said to be \textbf{$k$-repeat free} (or, interchangeably, \textbf{weak De-Bruijn of span $k$}) if every $k$-tuple appears \emph{at most} once as a substring in $w$. The set of length-$n$ $k$-repeat free sequences is denoted by 
	\[\hat{\cW}_k(n)\eqdef \mathset{w\in\Sigma^{n} ~:~ \forall u\in\Sigma^{k},\; \fr^k_w(u)\leq \frac{1}{n-k+1}}.\]
	For any $k$, we define the \textbf{$k$-repeat free system} (\textbf{weak De-Bruijn system}) as $\hat{\cW}_k=\bigcup_{n\in\N} \hat{\cW}_k(n)$. 
\end{definition}
Note that if $n=q^k+k-1$ then $\hat{\cW}_k(n)$ is exactly the set of all De-Bruijn sequences of span $k$. On the other hand, if $n> q^k +k-1$ then $\hat{\cW}_k(n)=\emptyset$ since there are more substrings than $k$-tuples.
This implies that for any fixed $k$ we have 
\[\ccap(\hat{\cW}_k)=\limsupup{n} \frac{1}{n}\log_q |\hat{\cW}_k(n)|=-\infty.\]
Therefore, a more natural question to ask is how the size $|\hat{\cW}_k(n)|$ behaves when $k$ and $n$ grow together. Namely, we are interested in the set $\hat{\cW}_k(n)$ where $k>\log (n-k+1)$ and is a function of $n$. We will calculate the capacity of a $k$-repeat free system for $k=\floorenv{a\log_q(n)}$ with $a>1$. Under this scenario, we will denote $\hat{\cW}_k(n)$ as $\cW_a(n)$ and $\hat{\cW}_k$ as $\cW_a$. That is,  $\cW_a(n) = \hat{\cW}_{\floorenv{a\log_q (n)}}(n)$ and $\cW_a = \bigcup_{n\in\N} \hat{\cW}_{\floorenv{a\log (n)}}(n)$.

The size $|\cW_a(n)|$ will be estimated by a probabilistic approach. Consider the uniform distribution over all length-$n$ sequences, then $|\cW_a(n)|=|\Sigma|^{n}\cdot \Pr(\cW_a(n))$. Then, the capacity in this case is given by 
\begin{align}
\label{eq:c1}
\ccap(\cW_a)=1+\limsupup{n}\frac{1}{n}\log_q\left( \Pr(\cW_a(n))\right).
\end{align}

Using standard union bound arguments, it is possible to show that for $a\geq 2$, $\ccap(\cW_a) =1$. However, in the following theorem we apply a different method which assures that this capacity result holds for all $a>1$. 
\begin{theorem}
	\label{th:cap}
	Let $\Sigma$ be a finite alphabet of size $q$ then for all $a>1$, $\ccap(\cW_a)=1$. 
\end{theorem}

We prove Theorem \ref{th:cap} using the (asymmetric) Lov\'asz local lemma which was first proved in \cite{ErdLov75} and is stated next as appears in \cite{AloSpe00}. 
\begin{lemma}[{\cite[Lemma 5.1.1]{AloSpe00}}]
\label{lem:LLL}
    Let $A_0,\dots,A_{m-1}$ be events in an arbitrary probability space. Let $G=(V,E)$ be a graph with $V=[m]$ such that for every $i\in [m]$, the event $A_i$ is mutually independent of all the events $\mathset{A_j ~:~ (i,j)\notin E}$. Suppose that there are real numbers $x_0,\dots,x_{m-1}$ such that $x_i\in [0,1]$ and that $\Pr(A_i)\leq x_i\prod_{(i,j)\in E} (1-x_j)$ for all $i\in [m]$. Then 
    \[\Pr\parenv{\bigcap_{i\in [m]} \overline{A_i}}\geq \prod_{i\in [m]} (1-x_i)\]
    where $\overline{A_i}$ us the complement of $A_i$.
\end{lemma}

Observe that Lemma \ref{lem:LLL} is useful especially when there is a small amount of dependencies between the events, or, when every event depends on a small number of other events.

\begin{IEEEproof}[Proof of Theorem \ref{th:cap}]
    For simplicity of notation we consider sequences of length $n+k$ instead of $n$. Let $w=(w_0,\dots,w_{n+k-1})\in\Sigma^{n+k}$ be a sequence of length $n+k$ in which every symbol is drawn i.i.d uniformly at random from $\Sigma$. 
    For $u=(u_0,u_1)\in [n]^2$, we denote by $I_u=\mathbb{1}_{w_{u_0+[k]}}(w_{u_1+[k]})$ the indicator function of the event that the $k$-tuples that start in positions $u_0$ and $u_1$ are identical. We are interested in a lower bound on 
    \[\Pr\parenv{\sum_{0\leq u_0<u_1 < n} I_{(u_0,u_1)}=0}=\Pr\parenv{\sum_{u\in \cI} I_{(u_0,u_1)}=0}\] 
    where $\cI:=\mathset{u=(u_0,u_1)\in [n]^2 ~:~ u_0\neq u_1}$. It is clear that if $u=(u_0,u_1),v=(v_0,v_1)\in \cI$ are such that all the $k$-tuples that start at locations $u_0,u_1,v_0,v_1$ do not overlap, then $I_u$ and $I_v$ are independent. Moreover, it is straightforward to show that $\Pr\parenv{I_u=1}=\frac{1}{q^k}= \frac{1}{n^{1+\epsilon}}$ for every $u\in \cI$. 
    
    We will now use Lemma \ref{lem:LLL} with 
    $\mathset{A_i ~:~ i\in [m]}= \mathset{(I_u=1) ~:~ u\in \cI}$, i.e., the set $[m]$ in the lemma corresponds to the set $\cI$ of size $\binom{n}{2}$, and the event $(I_u=1)$ corresponds to the set of all sequences for which $I_u=1$. For $u=(u_0,u_1),v=(v_0,v_1)$, we draw an edge $(u,v)\in \cI^2$ if at least one of $u_0+[k],u_1+[k]$ overlaps with $v_0+[k]$ or $v_1+[k]$. Thus, every $u\in \cI$ has at most $4kn$ neighbours. 
    
    We set the real numbers to be $x_u= \frac{1}{4nk}$ for every $u\in \cI$. 
    For $n$ large enough the condition of the lemma holds since 
    \[\prod_{(u,v)\in E} (1-x_v)= \parenv{1-\frac{1}{4nk}}^{4nk}\to e^{-1}\] 
    and for $n$ large enough 
    \[\Pr(I_u=1)=\frac{1}{n^{1+\epsilon}}\leq \frac{1}{4nk}(e^{-1}+o(1)).\]
    We obtain 
    \[\Pr\parenv{w\in \cW_a}=\Pr\parenv{\prod_{u\in \cI}(1-I_u)=1}\geq \prod_{u\in \cI} \parenv{1-\frac{1}{4nk}}.\]
    Since $\parenv{1-\frac{1}{4nk}}\leq 1$ and since $|\cI|\leq n^2$ we have 
    \[\Pr\parenv{w\in \cW_a}\geq \parenv{1-\frac{1}{4nk}}^{n^2}.\]
    Since $\parenv{1-\frac{1}{4nk}}^{n^2}\sim \exp \parenv{-\frac{n}{4a\log (n)}}$, we obtain that 
    \begin{align*}
        &\limsupup{n}\frac{1}{n+k}\log_q \parenv{\Pr\parenv{w\in \cW_a}} \\ 
    &=\limsupup{n}\frac{1}{n+k}\log_q \parenv{e^{-\frac{n}{4a\log_2 n}}}\\ 
    &=0
    \end{align*} 
    and the result follows.
\end{IEEEproof}

\section{Algorithm For $k=2\log (n)+2$} \label{sec:alg1}
%%%%%%%%%%%%%%%%%%%%%%%%%%%%%%%
In this section, we provide a coding algorithm for the binary weak De-Bruijn system with $k=2\log (n)+2$, where for simplicity we assume that $n$ is a power of 2. This will be the basic step towards an algorithm for the case $k=a\log (n)$ with $a>1$ that will be presented in Section~\ref{sec:alg2}. 

The input to the algorithm is a binary sequence $w\in\Sigma^{n-2}$, where in this section $\Sigma=\{0,1\}$. The output is a $k$-repeat-free sequence $\overline{w}\in\hat{\cW}_k(n)$. We first give a short overview of the algorithm, which is divided into two procedures: \emph{elimination} and \emph{expansion}. During the elimination phase, repeated $k$-tuples are deleted and the sequence is shortened. During the expansion phase, we append symbols to the sequence such that the $k$-repeat free constraint is remained and the sequence length reaches $n$ (notice that the encoding procedure must return a sequence of length $n$).

Given a sequence $w\in\Sigma^{n-2}$, first append $10^{1+\log(n)}$ to its end and append $0$ to its beginning. These appended strings serve as a markers that mark the information containing part of the sequence. Next, search for identical substrings of length $2\log(n)+2$. For every such occurrence, remove one of them (the first one) and encode at the beginning of the sequence $0$ followed by the indices of these two substrings. Note that such an operation reduces the length of the sequence by one, and therefore this procedure is guaranteed to terminate. Repeat this step until there are no more repeated $k$-tuples. Obviously, the resulting sequence will not contain repeated $k$-tuples, but it may be shorter than $n$. The second procedure takes this compressed sequence and decompresses it into a longer sequence such that the constraint is not violated. This is done in a straightforward manner- look for a $k$-tuple such that appending this $k$-tuple to the end of the sequence will not violate the $k$-repeat free constraints, and repeat this process until the sequence reaches length $n$. We show that it is always possible to find such a $k$-tuple. The output is the first $n$ bits of the decompressed sequence. 
Notice that in the second procedure we increase the length of the sequence by appending bits that do not contain any information. The reason that we appended the marker $10^{\log(n)+1}$ is to distinguish between the information symbols and between the symbols that we append at the expansion procedure. In order for this marker to be effective, it is necessary that it will be unique and hence, in addition to the elimination of repeated $k$-tuples, we will also eliminate appearances of the string $0^{\log(n)+1}$ (except for the one that we added during the encoding procedure). Before describing the algorithm in details, we present an example for the encoding procedure.

\begin{example}
		\label{ex:sec}
	Let $n=32$ $(k=2\log(32)+2=12)$ and  
	\[w=1111111111 1101011111 1111111111 \in\Sigma^{30}.\]
	At first, we append the marker $10^{\log(n)+1}=1000000$ to the end of the sequence and $0$ to the beginning of the sequence 
	\[\overline{w}= 0 1111111111 1101011111 1111111111 1000000 .\]
	Next, we look for repeated $k$-tuples ($k=12$). We note that the first $k$-tuple equals the $15$th, i.e., $\overline{w}_{[12]}=\overline{w}_{15+[12]}$. We eliminate the first $12$ bits and encode the locations by appending to the left $0$ and then the starting positions of the repeated $12$-tuples (in binary representation) $00000$ and $01111$. Hence the new sequence we obtain is 
	\[\overline{w}= 0\; 00000\; 01111\; 1010111111111111111 1000000.\]
	Next, we again look for repeated $12$-tuples and find that $\overline{w}_{15+[12]}=\overline{w}_{16+[12]}$. We again eliminate $\overline{w}_{15+[12]}$ and append $0\; 01111\; 10000$ to the left. We obtain 
	\[\overline{w}= 0\; 01111\; 10000\; 0\; 00000\; 01111\; 1010111 1000000.\]
	Now there are no more repeated $12$-tuples. Nevertheless, our marker $1000000$ appears twice in $\overline{w}$, once at the end as (as placed at the beginning of the encoding process), and also starting in position $6$, $\overline{w}_{6+[7]}=1000000$. In order to keep our marker unique, we eliminate the sequence $000000$ that appears in $\overline{w}_{7+[6]}$. We encode this elimination by appending the the left $1$ followed by the location of the sequence $000000$, i.e., we append $1\; 00111$ to the left. We obtain 
	\[\overline{w}= 1\; 00111\; 0\; 01111\; 10000\; 01111\; 1010111 1000000.\]
	Now there are no more repeated $12$-tuples and also the marker is unique. Notice that $|\overline{w}|=36>32$. Hence, we return only the first $32$ bits 
	\[\overline{w}=1\; 00111\; 0\; 01111\; 10000\; 01111\; 1010111 100.\]
\end{example}

In order to describe the algorithm explicitly, we need a few more notations. For an integer $i\in [n]$, we let $\bfb(i)$ be its binary representation using $\log(n)$ bits. Let $w\in\Sigma^n$ be any word. Recall that for $i\in\N$, $i\leq |w|$, $w_{[-i]}$ is the length-$i$ suffix of $w$, i.e., $w_{[-i]}=w_{|w|-i+[i]}$. Moreover, the support of $\fr^k_w$, $\Supp (\fr^k_w)$, is the set of all $k$-tuples that appear in $w$. For a word $w\in\Sigma^n$ and for $m\in\N$ we denote by $\Cr_m(w)$ the word of length $m$ created by repeatedly concatenating $w$ to itself and taking the length-$m$ prefix, i.e., $\Cr_m(w)=(w^{\N})_{[m]}$. 
We say that a sequence $w \in \Sigma^*$ is \textit{\textbf{$\ell$-zero-constrained}} if there are no all-zeros substrings of length $\ell$. We say that $(i,j)$ (where $i<j$) is a \textit{\textbf{$k$-identical window}} in $w$ if $w_{i + [k]} = w_{j + [k]}$. If $(i,j)$ is such that for any other $k$-identical window at $(i',j')$ in $w$, we have $j \leq j'$, we say that $(i,j)$ is a \textit{\textbf{primal $k$-identical window}}. The full details appear in Algorithm~\ref{alg:encoding1}.
\begin{algorithm}
	\caption{No-Identical Windows Encoding}\label{alg:encoding1}
	\begin{algorithmic}[1]
		\vspace{.1ex}
		\Require Sequence $w \in \Sigma^{n-2}$ 
		\Ensure Sequence $\overline{w} \in \hat{\cW}_k(n)$ with $k=2\log(n)+2$ \newline
		First procedure (elimination):
		\State Set $\overline{w} = 0\circ w \circ 1 \circ 0^{\log(n)+1} \in \Sigma^{n+\log(n)+1}$
		\While{$(i,j)$ is a $k$-identical windows in $\overline{w}$ or $\overline{w}_{[|\overline{w}|-1]}$ is not a $(\log(n)+1)$-zero-constrained (check the $1$st condition first)}
		\State \textbf{Case 1}: (there are identical length-$k$ windows in $\overline{w}$)
		\State \ \ \ Let $(i,j)$ be a primal $k$-identical window in $\overline{w}$
		\State \ \ \ Set $\overline{w}= \overline{w}_{[i]}\circ\overline{w}_{i+k+[|\overline{w}|-k-i]}$ (remove the first length-$k$ repeated window from $\overline{w}$)
		\State \ \ \ Set $\overline{w} = 0\circ \bfb(i)\circ\bfb(j)\circ\overline{w}$ (append $0\circ\bfb(i)\circ\bfb(j)$ to the left of $\overline{w}$)
		\State \textbf{Case 2}: ($\overline{w}_{[|\overline{w}|-1]}$ is not a $(\log(n)+1)$-zero-constrained)
		\State \ \ \ Let $i$ be the index of the $0^{\log(n)+1}$-window in $\overline{w}$
		\State \ \ \ Set $\overline{w} = \overline{w}_{[i]}\circ\overline{w}_{i+\log(n)+[|\overline{w}|-i-\log(n)]}$ (remove the $0^{\log(n)+1}$-window from $\overline{w}$)
		\State \ \ \ Set $\overline{w} = 1\circ \bfb(i)\circ\overline{w}$ (append $1\circ\bfb(i)$ to the left of $\overline{w}$)
		\EndWhile
		\If{ $|\overline{w}| \geq n$} 
		\State Return $\overline{w}_{[n]}$
		\EndIf \newline
		Second procedure (expansion):
		\While{$|\overline{w}| < n$}
		\State Set \[B=\Supp\parenv{\fr^{\log (n)}_{\overline{w}}}\bigcup \bigcup_{1\leq i\leq \log(n)-1}\Cr_{\log(n)}(\overline{w}_{[-i]}).\]
		\State Set $S =\Sigma^{\log(n)} \setminus B$ and find $u\in S$
		\State Set $\overline{w} =  \overline{w} \circ u$ (append $u$ to the right of $\overline{w}$)
		\EndWhile
		\State Return $\overline{w}_{[n]}$
	\end{algorithmic}
\end{algorithm}

Before we show the correctness of the Algorithm, we explain it more thoroughly. 
As mentioned previously, in Step $1$ we append the string $10^{\log (n)+1}$ to the right of the sequence in order to mark the end of the information sequence, and append $0$ to the left to mark the beginning of the information sequence. In order for $10^{\log (n)+1}$ to be unique and serve as a marker, we will eliminate all other appearances of $0^{\log(n)+1}$. Therefore, in Step $2$ we are searching for a violation of the $k$-repeat free constraint or another appearance of $0^{\log (n)+1}$. If there is a violation of the $k$-repeat free constraint, i.e., there is an $(i,j)$ $k$-identical window, then we eliminate the first appearance and append $0b(i)b(j)$ to the left of the sequence. Since the length of $b(i)$ is $\log (n)$, the length of the resulting sequence is shorter by $1$. When there are no more repeated $k$-tuples, we search for an appearance of $0^{\log (n)+1}$. If a substring $0^{\log (n)+1}$ appears in location $i<n$ then we eliminate it and append $1b(i)$ to the left. Notice that in this case the resulting sequence is of the same length. We repeat those steps until the obtained sequence is $k$-repeat free and does not contain the substring $0^{\log (n)+1}$ other the one at the end. Now there are two cases: the first is that the length of the obtained sequence is more than, or equals to $n$ (this could be the case since at the first step we add bits to the sequence, making it longer than $n$). In this case we return the first $n$ symbols. The second case is that the obtained sequence is shorter than $n$. In this case we need to extend the sequence to length $n$. This is done in Steps $15-19$, where we are searching for a $k$-tuple that appending it to the right of the sequence will not violate the $k$-repeat free constraint. Notice that in Step $16$ we define the set $B$ to be the set of all substrings of length $\log(n)$ that appear in $\overline{w}$ together with the set $\bigcup_{1\leq i\leq \log(n)-1}\Cr_{\log(n)}(\overline{w}_{[-i]})$. The latter is the set of all length $\log(n)$ sequences that do not necessarily appear in $\overline{w}$ but concatenating them to the right of $\overline{w}$ will violate the $k$-free repeat constraint. We repeat this process until the obtained sequence is of length at least $n$ and return the first $n$ symbols. In this part of the process, it is possible that an appended string will generate the sequence $0^{\log(n)+1}$. Nevertheless, the procedure will work as long as the first appearance of $0^{\log(n)+1}$ is at the end of the part that contains the information.

We now show the correctness of Algorithm~\ref{alg:encoding1}. Notice that the first while loop ends since after every iteration either the length of the word $\overline{w}$ decreases by one (case $1$) or its Hamming weight increases (case $2$). Moreover, in Step $12$, the word $\overline{w}$ has no identical length-$k$ windows and has no $0^{\log(n)+1}$-window besides the one at its end, i.e., the word $\overline{w}_{[|w|-1]}$ is $(\log (n)+1)$-zero-constrained. We start with the following lemma.
\begin{lemma}\label{lem1}
	In Step 12, the vector $\overline{w}$ ends with the sequence $1\circ 0^{\log(n)+1}$.
\end{lemma}
\begin{IEEEproof}
	For any iteration of the first while loop for which there are two identical windows of length $k$ in $\overline{w}$, let $i$ and $j$ be their indices, where $i<j$. We claim that the value of $i$ satisfies $i\leq |\overline{w}| -3\log(n)-2$ and thus the last $\log(n)+1$ bits of the vector $\overline{w}$ are not removed.  Assume in the contrary that $|\overline{w}| -3\log(n)-2< i < j$. Then, the length-$k$ window starting at position $i$ has a 1 in its $(|\overline{w}|-\log(n)-i)$-th position while the length-$k$ window starting at position $j$ has a 0 in this position, which is a contradiction. It is also readily verified that the sequence $1\circ 0^{\log(n)+1}$ cannot be removed as part of a removal of a $0^{\log(n)+1}$-window. 
\end{IEEEproof}

\begin{lemma}\label{lem:step12}
	If the condition in Step $12$ holds, then the returned sequence is of length $n$ and has no identical windows of length $k$.
\end{lemma}
\begin{IEEEproof}
	This lemma follows directly from Step $13$. Indeed, since the first while loop ended, the returned sequence does not contain repeated $k$-tuples. From Step $13$ it is clear that the returned sequence is of length $n$.
\end{IEEEproof}

\begin{lemma}
	For every iteration of the second while loop, the set $S$ in Step $17$ is not empty.
\end{lemma}
\begin{IEEEproof}
	Note that the size of the set $B$ is at most $(|\overline{w}| -\log(n)+1) +(\log(n)-1) = |\overline{w}|<n$ and hence $B\neq\Sigma^{\log(n)}$.
\end{IEEEproof}

\begin{lemma}\label{lem:appear_once}
	For every iteration of the second while loop, in Step $18$ the new vector $\overline{w}'= \overline{w}\circ u$ contains the sequence $u$ exactly once at its end. 
\end{lemma}
\begin{IEEEproof}
	According to the construction of the set $B$, the sequence $u$ can appear in $\overline{w}'= \overline{w}\circ u$ only as a substring starting at positing $j$, where $|\overline{w}| - \log(n) +1 \leq j\leq |\overline{w}|-1$. Assume in contrary that there exists a value $j$ such that $(\overline{w}')_{j+[\log (n)]} = (\overline{w}\circ u)_{j+[\log (n)]} =  u$. But this implies that $u\in\Cr_n(\overline{w}_{[-i]})$ for some $1\leq i \leq \log(n)-1$ which is a contradiction to the construction of the set $B$ in Step $16$. 
\end{IEEEproof}

Let $\overline{w}_0$ be the value of the vector $\overline{w}$ after Step $14$ and $n_0=|\overline{w}_0|$ is its length. Assume that there are $\ell$ iterations of the second while loop, so the value of the vector $\overline{w}$ after Step $19$ is given by 
$$\overline{w} = \overline{w}_0\circ u_1\circ u_2\circ \cdots \circ u_\ell,$$
where $u_1,u_2,\ldots,u_\ell$ are the vectors which were appended to the right of the vector $u$ at each iteration of the while loop. 
\begin{lemma}\label{lem:secondloop}
	For $1\leq i\leq \ell$, the vector $\overline{w}_i = \overline{w}_0\circ u_1\circ u_2\circ \cdots \circ  u_i$ has no identical length-$k$ windows.
\end{lemma}
\begin{IEEEproof}
	We prove the lemma's statement by induction on the values of $i$. For the base case, we start with $i=1$ and show that the vector $\overline{w}_1 = \overline{w}_0\circ u_1$ has no identical length-$k$ windows. 
	
	Assume in the contrary that $(i,j)$ is a $k$-identical window. We only need to consider the cases where at least one of these two windows overlaps with $u_1$. This implies that the length-$k$ window starting at position $j$ overlaps with $u_1$, that is, 
	$$n_0-k+1\leq j \leq n_0+\log(n)-k.$$ 
	In particular, the window $(\overline{w}_1)_{j+[k]}$ contains the $0^{\log(n)+1}$-window at the end of $\overline{w}_0$. If $ i \leq n_0-k$, then according to Lemma~\ref{lem1}, $(\overline{w}_1)_{i+[k]}$ does not contain a $0^{\log(n)+1}$-window, which is a contradiction. Thus we only need to consider the case $n_0-k\leq i<j \leq n+\log(n)-k$. However, this implies that $(\overline{w}_1)_{j+[k]}$ is periodic with period $0\leq j-i\leq \log(n)-1$ which is impossible since it contains the pattern $1\circ 0^{\log(n)}+1$.	
	
	Next we prove the statement for $\overline{w}_2 = \overline{w}_0\circ u_1\circ u_2$. According to the induction assumption we only need to consider values of $i$ and $j$ such that there is an overlap with $ u_2$. 
	Hence,
	\begin{align*}
	 j&\leq n_0+2\log(n)-k = n_0-2, \\
	j&\geq n_0+\log(n)-k+1=n_0-\log(n)-1.
	\end{align*}
	In particular, the window $(\overline{w}_2)_{j+[k]}$ contains $u_1$ as a substring. However, since 
	$(\overline{w}_2)_{j+[k]}= (\overline{w}_2)_{i+[k]}$ we get that the sequence $u_1$ appears one more time in $\overline{w}_0\circ u_1$, which is a contradiction to Lemma~\ref{lem:appear_once}.
	
	Next we assume that the lemma's statement holds for $\overline{w}_i$ and prove that it holds for $\overline{w}_{i+1}$, where $1\leq i<\ell$. According to the induction assumption we only need to consider values of $i$ and $j$ such that there is an overlap with $u_{i+1}$. 
	Hence,
	$$n_0+i\log(n)-k+1\leq j\leq n_0+(i+1)\log(n)-k.$$ 
	In particular, the window $(\overline{w}_{i+1})_{j+[k]}$ starting at index $j$ contains the sequence $u_i$. However, since 
	$(\overline{w}_{i+1})_{j+[k]}= (\overline{w}_{i+1})_{i+[k]}$ we get that the sequence $u_i$ appears one more time in $\overline{w}_i$, which is a contradiction to Lemma~\ref{lem:appear_once}.
\end{IEEEproof}

\begin{theorem}
	Algorithm~\ref{alg:encoding1} successfully returns a $k$-repeat  free sequence.
\end{theorem}
\begin{IEEEproof}
	In case the condition in Step $12$ holds then according to Lemma~\ref{lem:step12},  Algorithm~\ref{alg:encoding1} returns a sequence with no identical length-$k$ windows. Otherwise, this claim holds from Lemma~\ref{lem:secondloop}.
\end{IEEEproof}

Note that there may be two identical length-$k$ windows which intersect, i.e., $(i,j)$ is a $k$-identical window with $j-i<k$. In this case, Step $5$ in the algorithm suggests to remove the first length-$k$ repeated window. This will not cause any problem since if $(i,j)$ is such a $k$-identical window, then it implies that $w_{i+[k]}$ is a periodic sequence with period $j-i$ and as such can be obtained from  the remaining bits. Nevertheless, this should be taken into account in the decoding process that is described next.

The decoding procedure is relatively simple. Look first for the left most sequence of $1\circ 0^{\log(n)+1}$. According to Algorithm~\ref{alg:encoding1}, everything to the right of this sequence was added during the expansion procedure and hence it can be removed. If there is no such $1\circ0^{\log (n)}$ window, look for the right-most $1$. Since the algorithm returns a sequence which is longer by $2$ than the input sequence, the right-most $1$ (and the zeros following that $1$) is a part of the initial set-up of the algorithm and should be removed. Next, if the length of the obtained sequence is $(n-1)$ we look at the first bit. If this bit is $0$ then we eliminate it and return the obtained sequence. Otherwise, if the bit is $1$ then this implies that a substring $0^{\log(n)+1}$ was eliminated. We reconstruct it according the metadata that appears at the beginning of the sequence and we check again the first bit. We continue repeating this step until the first bit is $0$. If the length of the obtained sequence is less than $(n-1)$, we do the following. If the first symbol is $1$, let $i$ be the position indicated by the $(\overline{w})_{1+[\log (n)]}$, i.e., $\bfb(i)=(\overline{w})_{1+[\log (n)]}$. Delete the first $\log(n)+1$ bits and enter $0^{\log(n)+1}$ in the $i$th position. If the first symbol is $0$, let $i$ and $j$ be the positions indicated by $(\overline{w})_{1+[\log (n)]}$ and by $(\overline{w})_{1+\log (n)+[\log (n)]}$, respectively. Let $u=(\overline{w})_{j-1+[k]}$, delete the first $2\log(n)+1$ bits, and put $u$ in the $i$th position. Repeat this process until obtaining a sequence of length $n-1$. Notice that during the encoding process, a certain $\bfb(i)$ that was added may create $k$-repeated sequence which implies that it will be eliminated in a future step.  Since the decoding procedure is done sequentially, i.e., at each time we are reversing one encoding operation, and since each operation in the encoding process is reversible, the decoding process returns the correct sequence. Before writing the decoding algorithm we need the following notation. For a binary representation $\bfb(i)$ of the number $i$, we denote by $\bfb^{-1}(i)$ its decimal value. For a sequence $w\in\Sigma^n$ and for an integer $m<n$ we denote by $\elim{m}{w}$ the sequence obtained after eliminating the first $m$ symbols from $w$, i.e., if $w=111001010$ and $m=2$ then $\elim{2}{w}=1001010$.

\begin{algorithm}%\begin{small}	
	\caption{Decoding Process}\label{alg:decoding1}
	\begin{algorithmic}[1]
		\vspace{.1ex}
		\Require Sequence $\overline{w} \in \hat{\cW}_k(n)$ with $k=2\log(n)+2$
		\Ensure Sequence $w \in \Sigma^{n-2}$ 
		\If{ The substring $10^{\log (n)+1}$ appears in $\overline{w}$ } 
		\State  Set $w$ to be the sequence obtained after eliminating from $\overline{w}$ the left-most $10^{\log(n)+1}$ and all the bits following it. 
		\Else  
		\If{The substring $10^{\log (n)+1}$ does not appear in $\overline{w}$} 
		\State Set $w$ to be the sequence obtained after eliminating from $\overline{w}$ the right-most $1$ and all the zeros following it.
		\EndIf
		\EndIf
		\newline
		\While{$|w|\leq n-1$}
		\If{($w_0=1$)}
		\State Set $i=\bfb^{-1}(w_{1+[\log(n)]})$ ($i$ is the locations of the eliminated $0^{\log(n)+1}$)
		\State Set $w=\elim{\log(n)+1}{w}$ (eliminate the metadata)
		\State Set $w=w_{[i]}\circ 0^{\log(n)+1}\circ \elim{i}{w}$ (restore the eliminated $0^{\log(n)+1}$)
		\EndIf
		\newline
		\If{($w_0=0$ and $|w|<n-1$)}
		\State  Set $i,j$ as the locations of the $k$-repeats as follows
		\begin{align*}
		i&=\bfb^{-1}(w_{1+[\log(n)]}), \\
		j&=\bfb^{-1}(w_{1+\log(n)+[\log(n)]})
		\end{align*}
		\If{ ($j-i\geq k$) } 
		\State Set $u=w_{j-1+[k]}$ (set $u$ as the eliminated $k$-tuple)
		\State Set $w=\elim{(2\log(n)+1)}{w}$ (eliminate the metadata)
		\State Set $w=w_{[i]}\circ u\circ \elim{i}{w}$ (restore the eliminated $k$-tuple)
		\Else
		\If{($j-i<k$)}
		\State Set $u=w_{i+k-1+[j-i]}$ (the period of the eliminated $k$-tuple)
		\State Set $u= (u^k)_{[-k]}$ (generating a $k$-tuple using the period found earlier)
		\State Set $w=\elim{(2\log(n)+1)}{w}$ (eliminate the metadata)
		\State Set $w=w_{[i]}\circ u\circ \elim{i}{w}$ (restore the eliminated $k$-tuple)
		\EndIf
		\EndIf
		\EndIf
		\If{($w_0=0$ and $|w|=n-1$)}
		\State Set $w=\elim{1}{w}$ (eliminate the $0$ added to the left at the beginning of the encoding process)
		\State Return $w$ and stop
		\EndIf
		\EndWhile;
			\end{algorithmic}%\end{small}
	\end{algorithm}

In order to demonstrate the decoding algorithm, we apply the algorithm on Example \ref{ex:sec}. 

	\begin{example}[Continue Example \ref{ex:sec}]
		Recall that in Example \ref{ex:sec} the information sequence which is the input to the encoding process described in Algorithm \ref{alg:encoding1} is 
		\[ w=111111111111010111111111111111\in\Sigma^{30}.\]
		The output of the encoding process described in Algorithm \ref{alg:encoding1} is 
		\[	\overline{w}=1\; 00111\; 0\; 01111\; 10000\; 01111\; 1010111 100\in\Sigma^{32}.\]
		We now follow Algorithm \ref{alg:decoding1}. 
		At first, we notice that the substring $1000000$ does not appear in $\overline{w}$. Hence, we eliminate the right-most $1$ and the bits following it and set
		\[w=1\; \underset{1}{0} 011\underset{5}{1} \; \underset{6}{0} \; 01111\; 10000\; 01111\; 1010111\]
		where for convenience, we marked the locations of several bits (the numbers that appear beneath the bits). 
		We obtain that $|w|=29<31$. The first bit is $1$, i.e., $w_0=1$ which implies that we are performing Step $10$. We have $w_{1+[5]}=00111$ which implies that 
		$i=\bfb^{-1}(00111)=7$. In Step $11$ we set 
		\[w=\underset{0}{0}\; 01111\; \underset{6}{1} \underset{7}{0} 000\; 01111\; 1010111\] 
		and in Step $12$ we set
		\begin{align*}
		w&=w_{[7]}\circ 000000 \circ \elim{7}{w} \\
		&= \underset{0}{0}\; \underset{1}{0}1111\; \underset{6}{1} \; 0000\underset{11}{0}0 \; 0000\; 01111\; 1010\underset{26}{1}11.
		\end{align*}
		Again we obtain $|w|=29<31$ but now $w_0=0$. In Step $15$ we set $i,j$ as follows
		\begin{align*}
		i&=\bfb^{-1}(01111)=15, \\
		j&= \bfb^{-1}(10000)=16.
		\end{align*}
		Since $j-i=1<12$ we move to Step $22$ in which we set 
		\[u=w_{15+12-1+[1]}=w_{26}=1.\] 
		Next, we set 
		\[u=(u^{12})_{[-12]}=111111111111\] 
		and also
		\[w=00 \; 0000\; 01111\; 101\underset{14}{0}111.\]
		In Step $25$ we write
		\begin{align*}
		w&=w_{[15]} \circ 111111111111 \circ \elim{15}{w} \\
		&= \underset{0}{0}\underset{1}{0} \; 0000\; \underset{6}{0}111\underset{10}{1}\; 101\underset{14}{0} \; 1111111111\underset{25}{1}1 \; 11\underset{29}{1}.
		\end{align*}
		
		The obtained sequence $w$ is of length $|w|=30<31$. Therefore, we repeat the while loop in Step $8$. Again $w_0=0$ and $|w|<31$ therefore we obtain 
		\begin{align*}
		i&=\bfb^{-1}(00000)=0, \\
		j&= \bfb^{-1}(01111)=15.
		\end{align*}
		Since now $j-i=15>12$ we move to step $17$.
		We set 
		\[u=w_{15-1+[12]}=011111111111\] 
		and set 
		\[w=1010 \; 111111111111\; 111.\]
		After Step $19$ we obtain 
		\begin{align*}
		w&= w_{[0]}\circ u \circ \elim{0}{w} \\
		&= \underset{0}{0}1111111111\underset{11}{1} \; 1010 \; 111111111111\; 111.
		\end{align*}
		The obtained sequence is of length $|w|=31$. We again repeat the loop in Step $8$ but since $w_0=0$ and $|w|=31$ we jump to Step $29$. We set 
		\[w=\elim{1}{w}=11111111111 \; 1010 \; 111111111111\; 111\in\Sigma^{30},\]
		return $w$ and stop. We decoded the word correctly.
	\end{example}

We now give another example of the entire encoding and decoding process.
\begin{example}
	Let $n=32$ ($k=2\log(32)+2=12$) and 
	\[w= 100100110110010010011011100110\in\Sigma^{30}.\] 
	The first step of the algorithm appends $1000000$ to the end of $w$ and  append $0$ to the beginning to obtain 
	\[\overline{w}=0100100110110010010011011100110\; 1000000.\]
	We now look for identical windows of length $12$. We see that $(\overline{w})_{[12]}=(\overline{w})_{13+[12]}$, i.e., $(0,13)$ is a $k$-identical window. We eliminate the first $12$ bits and we append $0\circ\bfb(0)\bfb(13)=00000001101$ to the left of $\overline{w}$. Hence,
	\[\overline{w}=00000001101\; 0010010011011100110\; 1000000.\]
	There are no more identical length-$k$ windows in $\overline{w}$, but the pattern $000000$ appears in $\overline{w}$ in the $0$th position. Thus, we eliminate the pattern and append $1\bfb(0)=100000$ to the left, which yields the sequence
	\[\overline{w}=100000\; 01101\; 0010010011011100110\; 1000000.\]
	Again, there is a sequence of $6$ zeros starting in position $1$ so we delete this pattern and append $1\bfb(1)=100001$ to the left, so we get that 
	\[\overline{w}= 100001\;1\;1101\; 0010010011011100110\; 1000000.\]
	Now $\overline{w}$ has no identical windows of length $k$ and no $0^{\log(n)+1}$ except the one at the end. Moreover, $|\overline{w}|\geq 32$ hence the algorithm output is 
	\[\overline{w}=100001\;1\;1101\; 0010010011011100110\; 10.\] 
	
	We now start the decoding process in order to retrieve $w$ from $\overline{w}$. First, we look for the left most $1000000$ substring in $\overline{w}$. Since there is no such sequence, we look for the right-most $1$ and we know that this bit with all the following zeros were added in the set-up. That is, the last $10$ are not part of $w$. We eliminate those bits and we obtain 
	\[w=100001\;1\;1101\; 0010010011011100110\in\Sigma^{30}.\] 
	Since $w\in\Sigma^{30}$ we know that there was only one identical pair of length-$k$ windows. The first bit in $w$ is $1$. Thus, we have $i=\bfb^{-1}(00001)=1$. We eliminate the first $6$ bits and insert $6$ zeros in the first position, 
	\[w=1\;000000\;1101\; 0010010011011100110.\]
	Again, the first bit is $1$ so the next $5$ bits indicate the position of the $0$. We eliminate the first $6$ bits and enter $000000$ in the $0$th position to get the word
	\[w=000000\;0\;1101\; 0010010011011100110.\]
	We are now having $0$ for the first bit and the next $10$ bits indicate two positions, $i=0$, $j=13$. We denote 
	\[u=(w)_{12+[12]}=010010011011.\]
	We now eliminate the first $11$ bits and put $u$ in the $i$th position and obtain 
	\[w=010010011011\;0010010011011100110.\]
	Since $w\in\Sigma^{31}$ and $w_0=0$ we eliminate the first $0$, return 
	\[ w=10010011011\;0010010011011100110\]
	and stop.
\end{example}

\section{Algorithm For $k=a\log (n)$ with $1<a < 2$} \label{sec:alg2}
%%%%%%%%%%%%%%%%%%%%%%%%%%%%%%%
In this section, we consider the case of $k=a\log(n)$ where $1<a<2$. Similarly to Section~\ref{sec:alg1}, our coding scheme consists of two basic procedures: elimination and expansion. Both procedures are similar to the procedures for the case $a\geq 2$ and enjoy the same intuition. Nevertheless, there are several differences. For the elimination phase, we compress an input sequence into an output sequence of length at most $n$. At every step of the compression, we remove repeated $k$-tuples so that at the end of this step, the output sequence is $k$-repeat free. The elimination process relies on an encoding procedure which is very similar to \cite{GabMil18}. The main differences from the case $a\geq 2$ are as follows. At first, we encode the input string to a new string that does not contain the substring $0^{2\log\log (n)}$. The reason behind this encoding will become clear later. Then, when we see an $(i,j)$ $k$-identical window, we eliminate the repeated substring that starts at position $j$. Notice that in our case (of $a<2$), the eliminated substring is of length $a\log(n)<2\log(n)$. Therefore, it will not be possible to append to the left the same meta-data that we appended in the case $a\geq 2$ since the string $0b(i)b(j)$ is longer than $a\log(n)$. Therefore, we need to find a better way to append the meta-data to our sequence, i.e., to use a shorter meta-data. This meta-data will no longer be appended to the left of the string, but it will replace the eliminated $k$-tuple that starts at location $j$. Thus, the location of the added meta-data implies the location of elimination. In this way, the added meta-data should contain information regarding the location $i$ (the location of the repetition) and also it should be distinguished from the rest of the information bits. This is done as follows. First, we encode the location $i$ to a codeword $f(i)$ that does not contain the sequence $0^{2\log \log (n)}$ using a constrained code. Next, we replace the eliminated $k$-tuple with the word $(1,0^{2\log\log (n)},1,f(i),1)$. We will show that this word is shorter than $k$ and thus the elimination process terminates. The substring $0^{2\log\log (n)}$ serves as a marker for a location of eliminated $k$-tuple, and $f(i)$ serves as a pointer to the location of the repeated substring. In order for this marker to be effective, the input sequence should not contain any substrings of the form $0^{2\log\log (n)}$.
Throughout this section we assume for simplicity that $\log (n)$ and $\log\log (n)$ are integers. Taking $\floorenv{\log (n)},\floorenv{\log\log (n)}$ will not affect the results.

The expansion phase is the primary difference between the approach outlined here and \cite{GabMil18}. The idea behind the expansion phase is to concatenate a zero-constrained De-Bruijn sequence, which we refer to as $v \in \Sigma^*$, with our compressed sequence, and then insert within $v$, all-zeros markers of length $4 \log \log (n)$. These markers will be used to distinguish (or to make different) the length-$k$ windows between $v$ and the compressed sequence. We will explain these ideas in more detail in what follows.
 
For $m\in \N$, let $S_m(n)$ denote the set of all sequences of length $n$ which are $(2\log\log (m))$-zero constrained, i.e., 
\[S_m(n)=\mathset{u\in\Sigma^n ~:~ \fr_u^{2\log\log (m)}(0\dots 0)=0}.\]
Note that $S_m=\bigcup_{n\in \N}S_m(n)$ is the $(0,2\log\log (m))$-RLL constrained system. It is well known (see, for example, \cite{MarRotSie2001}) that $\lim_{m\to\infty} \ccap (S_m)=1$. Moreover, the function $\log |S_m(n)|$ is subadditive in $n$ which implies, by Fekete's lemma, that the capacity of $S_m$ 
is obtained by $\inf_{n\in\N} \frac{1}{n}\log |S_m(n)|$. Therefore, there exists a large enough $n$ such that $\abs{S_{2\log\log (n)}(\log(n)+1)}\geq n$ (choose $n$ such that $\ccap(S_{2\log\log (n)})$ is close to $1$). 
Let $f : [n] \to \Sigma^{ \log (n) + 1}$ be a bijection from $[n]$ to $S_{2\log\log (n)}(\log (n) +1)$, i.e., the image of $f$ lies in the set of all $(2 \log \log (n))$-zero-constrained sequences. 

The elimination encoder $\cE_{el}$, described in Algorithm~\ref{alg:encoding} below, takes as input a sequence $w \in \Sigma^{n- (4\log \log (n) +3)}$, where $w$ is $(2 \log \log |w|)$-zero-constrained. The output of $\cE_{el}$ is a sequence $\overline{w}$ of length at most $n-(4 \log \log (n) + 3)$ that is $(2 \log \log |w|)$-zero-constrained and does not contain any repeated windows of length $k'=\log (n) + 2 \log \log (n) + 5$.

\begin{algorithm}[H]%\begin{small}	
	\caption{Elimination Encoder, $\cE_{el}$}\label{alg:encoding}
	\begin{algorithmic}[1]
		\vspace{.1ex}
		\State Set $\overline{w} = w$
		%	\State Set $end = n-1$
		\While{there are identical length-$k'$ windows in $\overline{w}$}
		\State Suppose $(i,j)$ is a primal $k'$-identical window in $\overline{w}$
		\State Remove the substring of length $k'$ starting at position $j$ and replace it with the sequence $(1,0^{2 \log \log (n)},1,f(i),1)$, so that
		\begin{align*}
		&\overline{w} = \\ \nonumber
		&\overline{w}_{[j]} \circ (1, 0^{2 \log \log (n)}, 1, f(i), 1) \circ \overline{w}_{\{j+k', j+k'+1,\ldots,|\overline{w}|-1\}} 
		\end{align*}
		\EndWhile
		\State Return $\overline{w}$
	\end{algorithmic}%\end{small}
\end{algorithm}

Note that since at Step 4 we replace substrings of length $k'$ with substrings of length $k'-1 = \log (n) + 2\log \log (n) + 4$, so that each time Step 4 is executed, the length of $\overline{w}$ is decremented by one. We have the following result, which follows from \cite{GabMil18}.

\begin{lemma}
	\label{lem:old} 
	(c.f., Claim10, \cite{GabMil18}) The sequence $\overline{w}$ has no repeated $k'$-windows and $\overline{w}$ can be recovered from $w$.  
\end{lemma}

In the following, let $k'=\log (n) + 2 \log \log (n) + 5$. For simplicity of calculations, we assume that $k'$ is a prime number, and we later show that we can relax this assumption since the result may be generalized to non prime numbers using similar techniques. For our construction, we require the use of Lyndon words and necklaces. For a word $w$, we say that $w$ is a \emph{Lyndon} word if $w$ is (strictly) smaller (with respect to the lexicographic order) than all of its rotations. A \emph{necklace} of length $k$ is an equivalence class of sequences of length $k$. Two sequences $w,u$ are equivalent (or, in the same necklace) if and only if they are equivalent under rotation, i.e., there exists $\ell$ such that  $(w_0,w_1,\dots,w_{k-1})=(u_{\ell},u_{\ell+1},\dots,u_{k-1},u_{0},\dots, u_{\ell-1})$. The next lemma follows from a well-known result on generating De-Bruijn sequences from Lyndon words \cite{FreKes86,Mor04}.

\begin{lemma}
	\label{lem:restrictDB} 
	The lexicographic concatenation of Lyndon words of length $k'$ which are greater (with respect to the lexicographic order) than or equal to the string 
	$$\parenv{(0^{2\log \log (n) - 1}\circ 1)^{k'}}_{[k']} $$
	generates a sequence of length greater than $n$ which does not contain any repeated windows of length $k'$ and also is $(4 \log \log (n))$-zero-constrained.
\end{lemma}

Before proving the previous lemma, we provide an example, which illustrates the idea behind the construction.

\begin{example} Suppose $k' = 5$. Then the Lyndon words of length $k'$ are:
\begin{align*}
&(0,0,0,0,0), (0,0,0,0,1), (0,0,0,1,1), (0,0,1,0,1), \\ 
&(0,0,1,1,1), (0,1,0,1,1), (0,1,1,1,1), (1,1,1,1,1).
\end{align*}
Concatenating these words together produces the De-Bruijn sequence 
\begin{align*} 
(0^5,1,0^3,1,1,0,0,1,0,1,0,0,1^3,0,1,0,1,1,0,1^5) 
\end{align*}
of length $32$. The key property to notice here is that the length of the runs of zeros is smaller towards the end of the sequence than at the beginning. For example, the longest run of zeros (of length $5$) appears in the first position and the last $8$ bits of the sequence contains only a single zero.
\end{example}

We now turn to the proof of Lemma~\ref{lem:restrictDB}.
\begin{IEEEproof} 
	It was established that the lexicographic concatenation of Lyndon words generates a De-Bruijn sequence \cite{FreKes86,Mor04}, and so it follows that our approach does not have any repeated windows of length $k'$. Let $w$ be the string which results by concatenating Lyndon words greater than or equal to $\parenv{(0^{2\log \log (n) - 1}\circ 1)^{k'}}_{[k']}$ as stated in the lemma. We now show $w$ is $(4 \log \log (n))$-zero-constrained, which implies the statement in the lemma.
	
	First, we recall a simple procedure from \cite{RusSavWan1992} which generates all Lyndon words of length $k'$. Let $\gamma : \{0,1\}^{k'} \to \{0,1\}^{k'}$ be such that given $x \in \Sigma^{k'}$, $\gamma(x) = y=(y_0, y_2, \ldots, y_{k'-1})$ where $y=\parenv{(x_{[j]}\circ 1)^{\N}}_{[k']}$ where $j$ is the largest index such that $x_{\mathset{j,j+1,\dots, (k'-1)}} = (011\dots 1)$. 
	Let 
	$$(0,0,\ldots,0), \gamma(0,0,\ldots,0), \gamma(\gamma(0,0,\ldots,0)),\dots$$ 
	be a sequence of sequences, and let $V$ denote the result of removing non-necklaces from this sequence. It is known that $V$ is a lexicographic (increasing) sequence of necklaces \cite{RusSavWan1992}. The string $w$ (mentioned two paragraphs above) is the result of concatenating the sequences (in order) from $V$.
	
	We show that if any $x \in V$ is $(2 \log \log (n) )$-zero-constrained, then the longest run of zeros in $\gamma(x)$ is $2\log \log (n) -1$, which implies that $w$ does not have any runs of length $4 \log \log (n)$. Assume in the contrary that it does not hold, so that $\gamma(x)$ contains an all-zero substring of length $2 \log \log (n)$. Let $j$ be the largest index that $x_{\mathset{j,j+1,\dots, (k'-1)}} = (011\dots 1)$ . Then according to the procedure from the previous paragraph, the all-zero substring of length $2 \log \log (n)$ occurs after index $j$ in $\gamma(x)$, since $x_{[j]}=\gamma(x)_{[j]}$ and $\gamma(x)_{j} = 1$. However, this is also not possible since $\gamma(x)_{[k']} $ comprises of repeated concatenations of $x_{[j]}\circ 1$, and so we arrive at a contradiction to the assumption that $x$ does not contain the all-zeros substring of length $2 \log \log (n)$. 
	
	We have left to show that $|w| > n$. To see this, note that since $k'$ is a prime, we can bound the length of the $|w|$ as follows. 
	\begin{align*}
	|w|& \stackrel{(a)}{\geq} \parenv{\frac{2^{k'}-2}{k'} - 2^{k' - 2\log \log (n)} }k'\\
	& \stackrel{(b)}{=} \parenv{\frac{n \cdot (\log (n))^2 \cdot 2^5 - 2}{\log (n) + 2 \log \log (n) + 5} - \frac{n (\log (n))^2 \cdot 2^5}{(\log (n))^2}}\cdot k' \\
	& = n2^5\parenv{(\log (n))^2-\log (n)-2\log\log (n)-5} -2 \\
	&\geq n,
	\end{align*}
	where $(a)$ follows since there are exactly $\frac{2^{k'}-2}{k'}$ necklaces of length $k'$ and there are at most $2^{k'-2\log\log (n)}$ words which are smaller than $\parenv{(0^{2\log \log (n) - 1}\circ 1)^{k'}}_{[k']}$, $(b)$ follows by plugging in the value of $k'$ and the last inequality holds for large enough $n$.
\end{IEEEproof}

\begin{remark}
	Note that the assumption that $k'$ is prime affects only the calculation of $|w|$. For a non prime $k'$, the calculation of $|w|$ is more involved (the expression for the number of necklaces is $ \frac{1}{k'}\sum_{d|k'}\mu(d)2^{\frac{k'}{d}}$ where $\mu$ is the m\"{o}bius function and the summation is over all divisors of $k'$). This results in the desired inequality for larger values of $n$. 
\end{remark}

Let $v'$ be the string of length at least $n$ generated from Lemma~\ref{lem:restrictDB}. Let $v$ be the result of inserting the all-zeros substring of length $4 \log \log (n)$ periodically into $v'$ as follows:

\begin{align}
\label{eq:d}
v= \Big ( v'_{[k']} \; 10^{4 \log \log (n)}1 v'_{k' + [k']}\; &10^{4 \log \log (n)}1\;\dots \\ \nonumber 
&v'_{\frac{|v'|}{k'} \cdot (k'-1) + [k']}   \Big ).
\end{align}

We have the following lemma. 

\begin{lemma}
	\label{lem:noidwindows} 
	Let 
	$$ \hat{w}=\Big ( \overline{w},1,0^{4 \log \log (n) +1}, 1, v \Big )_{[n]} $$
	be the substring of length $n$ which results by concatenating $\overline{w}$ and $v$. Then $\hat{w}$ does not contain any repeated windows of length $k=\log (n) + 10 \log \log (n) + 10=k'+8 \log \log (n) + 5$. 
\end{lemma}

\begin{IEEEproof} 
	Suppose, on the contrary, that there is a repeated $k$-window at $(i,j)$. The proof is done on a case-by-case basis and we show that for all options of $j>i$, $\hat{w}_{i + [k]} \neq \hat{w}_{j + [k]}$.
	
	If 
	\[|\overline{w}| + 4 \log \log (n) + 2 - k \leq i \leq |\overline{w}|+1\] 
	or 
	\[|\overline{w}| + 4 \log \log (n) + 2 - k \leq j \leq |\overline{w}|+1,\] 
	then $\hat{w}_{i + [k]} \neq \hat{w}_{j + [k]}$ since the all-zeros substring of length $4 \log \log (n) + 1$ appears only once in $\hat{w}$.
	
	If $j \leq |\overline{w}| + 4 \log \log (n) + 3 -k$, then the result follows immediately from Lemma~\ref{lem:old}.
	
	If $i > |\overline{w}|+1$, then we know that both $\hat{w}_{i+[k]}$ and $\hat{w}_{j + [k]}$ each contain the substring $(1,0^{4 \log \log (n)},1)$. Suppose for now that there is only one occurrence of $(1,0^{4 \log \log (n)},1)$ in $\hat{w}_{i + [k]}$. From (\ref{eq:d}), we know we can write:
	\begin{align*}
	\hat{w}_{i+[k]} = \Big ( \hat{w}^{(i,1)}, 1, 0^{4 \log \log (n)},1, \hat{w}^{(i,2)} \Big ).
	\end{align*}
	
	If $|\hat{w}^{(i,2)}| \geq 4 \log \log (n) + 3$, then from (\ref{eq:d}), we can recover a substring of $v'$ of length $k-4 \log \log (n) - 2$ by deleting the substring $(1, 0^{4 \log \log (n)}, 1)$ from $\hat{w}_{i+[k]}$. Otherwise if $|\hat{w}^{(i,2)}| = t < 4 \log \log (n) + 3$, then we can recover a substring of $v'$ of length 
	\[k-4 \log \log (n) - 2 -(4 \log \log (n) + 3 - t)=k' + t\] 
	by first deleting  the substring $(1, 0^{4 \log \log (n)}, 1)$ from $\hat{w}_{i+[k]}$ followed by deleting the first $4 \log \log (n) + 3 - t$ bits of the resulting string. The only case left to consider is where $\hat{w}_{i+[k]}$ contains two occurrences of the substring $(1,0^{4 \log \log (n)},1)$. Suppose the first occurrence of the substring  $(1,0^{4 \log \log (n)},1)$ appears in position $\ell$ where it is clear from (\ref{eq:d}) that $\ell \in \{0,1\}$. If $\ell = 1$, then we remove the first $4 \log \log (n) + 3$ bits from $\hat{w}^{(i,2)}$ followed by the last $4 \log \log (n) + 2$ bits. Otherwise, if $\ell=0$ we remove the first $4 \log \log (n) + 2$ bits from $\hat{w}^{(i,2)}$ followed by the last $4 \log \log (n) + 3$ bits to obtain a substring of $v'$ of length $k'$ from $\hat{w}_{i+[k]}$. 
	
	From the previous paragraph, we know we can recover distinct substrings of length at least $k'$ from $v'$ in $\hat{w}_{i + [k]}$ and $\hat{w}_{j + [k]}$ provided $i  > |\overline{w}|+1$. Since these substrings are unique from Lemma~\ref{lem:restrictDB}, it follows that $\hat{w}_{i + [k]} \neq \hat{w}_{j + [k]}$.
	
	We have left to consider the case where $j > |\overline{w}|+1$ and $i < |\overline{w}| + 4 \log \log (n) + 2 - k$. In this case, there are three possibilities for $\hat{w}_{i + [k]}$: a) $\hat{w}_{i + [k]}$ ends with the substring $0^{4 \log \log (n) + 1}$, b) $\hat{w}_{i + [k]}$ ends with the substring $0^{4 \log \log (n)}$, or c) $\hat{w}_{i + [k]}$ does not contain the substring $0^{4 \log \log (n)}$. If a) holds, then clearly $\hat{w}_{i + [k]} \neq w_{j + [k]}$, since by assumption $j > |\overline{w}|+1$ and $0^{4 \log \log (n) +1}$ only appears once in $\hat{w}$. If b) holds, then from (\ref{eq:d}), $\hat{w}_{j + [k]}$ contains two occurrences $0^{4 \log \log (n)}$, and $\hat{w}_{i + [k]}$ only has one occurrence so that $\hat{w}_{i + [k]} \neq \hat{w}_{j + [k]}$. Finally, if c) holds, then $\hat{w}_{i + [k]}$ does not contain the substring $0^{4 \log \log (n)}$ but $\hat{w}_{j + [k]}$ does and so $\hat{w}_{i + [k]} \neq \hat{w}_{j + [k]}$ in this case as well. 
\end{IEEEproof}

We now present our main result, which follows from the previous discussion.

\begin{theorem} 
	There exists a rate-$1$ polynomial-time encoder which generates sequences with no-identical $k$-windows for any $k > a \log (n)$ where $a > 1$.
\end{theorem}

\begin{IEEEproof} 
	The fact that our algorithm has polynomial-time encode complexity follows from the observation that $\cE_{el}$ runs in polynomial time along with the fact that generating a lexicographic ordering of Lyndon words can be accomplished in time at most $\cO(2^{k'})$ which is polynomial in $n$. Suppose $\hat{w}=( \overline{w},1,0^{4 \log \log (n) +1}, 1, v )_n$ is a codeword from Lemma~\ref{lem:noidwindows}. Then to recover $w$ from $\hat{w}$, we simply remove the suffix $(1,0^{4 \log \log (n) +1}, 1, v)$ from $\hat{w}$, which is the first suffix of $\hat{w}$ that begins with the substring $(1,0^{4 \log \log (n) +1}, 1)$, to recover $\overline{w}$. The result then follows immediately from Lemma~\ref{lem:old} since $w$ can be recovered from $\overline{w}$.
	
	Next, we verify the statement on the rate. From Claim~7 in \cite{GabMil18}, we have that there are at least
	$$\Bigg ( \frac{n}{4} \cdot \Big (1 - \frac{\log (n)}{(\log (n))^2} \Big ) \Bigg )^{\lfloor \frac{n - (4 \log \log (n)  + 3)}{\log (n) } \rfloor }, $$
	possible input sequences for Algorithm~2 since we can divide up the input sequence of length $n - (4 \log \log (n) + 3)$ into blocks of length $\log (n)$ that begin and end with the symbol $1$, and then constrain each block to have runs of zeros of length at most $2 \log \log (n) -1$. Then,
	$$ \lim_{n \to \infty} \frac{1}{n} \log \Bigg ( \frac{n}{4} \cdot \Big (1 - \frac{\log (n)}{(\log (n))^2} \Big ) \Bigg )^{\lfloor \frac{n - (4 \log \log (n)  + 3)}{\log (n) } \rfloor } = 1,$$
	which completes the proof. 
\end{IEEEproof}

\section{$k$-Repeat Free Sequences With Combinatorial Constraints}\label{sec:comb}
%%%%%%%%%%%%%%%%%%%%%%%%%%%%%%%%%%%%%
In this section we study the combination of $k$-repeat free sequences and combinatorial constraints. As mentioned previously, the number of De-Bruijn sequences of span $k$ can be calculated using several combinatorial methods such as the doubling process, the BEST theorem, and using shift registers. Unfortunately, calculating the exact number of De-Bruijn sequences which also satisfy other constraints is not an easy problem \cite{KupVar16}. Here, we calculate the capacity of $k$-repeat free systems with local constraints.
For convenience, throughout this section we restrict $\Sigma$ to the binary alphabet but the same method can be used for larger alphabets.

Before stating the main result of this section, we remind the reader some known definitions and basic results on constrained systems. We follow the lines of \cite{MarRotSie2001}.
Let $G=(V,E,L)$ be a labeled (directed) graph where $V$ is the set of vertices, $E$ is the set of edges and $L:E\to \Sigma$ is a labeling of the edges. We say that a graph $G$ is \emph{deterministic} if from every vertex, the outgoing edges have different labels. For each graph $G$, we denote by $A_G$ the adjacency matrix of $G$. The adjacency matrix is a $|V|\times |V|$ matrix such that the $u,v$ entry of $A_G$ is the number of edges which start at the vertex $u$ and end at $v$.

A constrained system $S\subseteq \Sigma^{\N}$ is the set of all words obtained by reading the labels of paths in a  labeled directed graph. If $S$ is obtained by a graph $G$, we say that $G$ presents the system $S$ (or $G$ is a presentation of $S$). A constrained system which is presented by a graph $G$ is said to be \emph{irreducible} if $G$ is strongly connected. For a system $S$, we denote by $\cB_n(S)$ the set of all length-$n$ blocks that appear in words in $S$. The language of $S$ is denoted by $\cB(S)=\bigcup_{n\in\N} \cB_n(S)$. 
It is well known that every constrained system can be presented by a deterministic graph \cite[Prop. 2.2]{MarRotSie2001}. Therefore, we will assume that all presentations are deterministic. The capacity of a constrained system $S$ is defined as $\ccap(S)=\limsupup{n} \frac{1}{n}\log_2 |\cB_n(S)|$. The adjacency matrix is highly related to the capacity of the system. If $S$ is irreducible, the Perron-Frobenius theorem states that $A_G$ has a largest, real, simple eigenvalue $\lambda$, with strictly positive left and right eigenvectors. If $S$ is irreducible, it is well known that $\ccap(S)=\log_2 \lambda$~\cite[Th. 3.4]{MarRotSie2001}.

In this section we are interested in constrained systems which are also repeat free. In other words, if $S$ is a constrained system, we are interested in the following set of words.
\begin{definition}
	Let $S$ be an irreducible deterministic constrained system with language $\cB(S)$ and let $\hat{\cW}_k(n)$ denote the $k$-repeat free sequences. The $(S,k)$-repeat free system with is defined by the following sets,
	\[\cX_{S,k}(n)=\mathset{w\in\Sigma^n ~:~ w\in\hat{\cW}_k(n)\cap \cB_n(S)}.\]
	We define the system $\cX_{S,k}=\bigcup_{n\in\N}\cX_{S,k}(n)$.
\end{definition}

We are interested in the capacity of the system $\ccap(\cX_{S,k})$, where $k=k(n)$ grows with $n$. In order to estimate this capacity we need the following useful characterization of the capacity of a constrained system given by Markov chains. For a graph $G=(V,E)$, a Markov chain is given by a transition probability matrix $P\in [0,1]^{|V|\times |V|}$ such that $P\cdot \1=\1$, where $\1$ is the all ones vector. For an edge $e\in E$, we denote by $e^b$ the starting vertex of $e$ and by $e^t$ the terminal vertex, i.e., if $e=(u,v)\in E$ then $e^b=u$ and $e^t=v$. Thus, from a vertex $u$, the $(u,v)$ entry of $P$ corresponds to the transition probability from vertex $u$ to vertex $v$. If for every $u,v\in V$ there exists $n\in \N$ such that $(P^n)_{u,v}>0$ then we say that $P$ is \emph{irreducible}. For an irreducible Markov chain $P$, there is a unique positive stationary vector $\mu^T$ such that $\mu^T P=\mu^T$. For a Markov chain $P$ on a graph $G=(V,E)$ with stationary distribution $\mu$, the entropy of the Markov chain is defined as 
\[H(P)=-\sum_{u\in V}\mu_u\sum_{(u,v)\in E}P_{u,v}\log_2 P_{u,v}.\]
We may now state the known relation between the capacity and Markov chains~\cite[Th. 3.23]{MarRotSie2001}.
\begin{theorem}\cite[Th. 3.23]{MarRotSie2001}
	\label{th:capmarkov}
	Let $S$ be an irreducible constrained system presented by $G$ with Perron eigenvalue $\lambda$. Then
	\[\sup_P H(P)=\log_2 \lambda=\ccap(S),\]
	where the supremum is taken over all Markov chains on $G$.
\end{theorem}

In order to use Theorem~\ref{th:capmarkov}, we need to find the entries of $P$ that maximize the entropy. Note that although in Theorem~\ref{th:capmarkov} we take supremum over all Markov chains, the set on which we take the supremum is a compact set, which means that the supremum is in fact a maximum. Moreover, a closer look on the proof of Theorem~\ref{th:capmarkov} (as in \cite[Th. 3.23]{MarRotSie2001}), reveals exactly the maximizing transition probabilities and the corresponding stationary vector. Indeed, let $A_G$ be the adjacency matrix of $G$ and denote by $\eta^T,\nu$ the normalized left and right eigenvectors of the Perron eigenvalue $\lambda$ such that $\eta^T \nu=\1$. Then, the Markov chain which maximizes the entropy is given by 
\[P_{u,v}=\frac{(A_G)_{u,v} \nu_v}{\lambda \nu_u},\]
and the corresponding stationary vector is given by $\mu_u=\eta_u^T \nu_u$. This means that all the edges from $u$ to $v$ are prescribed with the same probability which is $\frac{\rfrac{\nu_u}{\lambda \nu_v}}{(A_G)_{u,v}}$.
In the next lemma we show that in a constrained system, the probability of two $k$-tuples to be identical is upper bounded by a constant times $\lambda^{-k}$.
\begin{lemma}
	\label{lem:shiftinv}
	Let $S$ be an irreducible constrained system presented by $G=(V,E)$ with an entropy maximizing Markov chain $P$. Let $x\in\Sigma^{\N}$ be a sequence obtained by reading the labels of a path evolving according to $P$ with the initial state chosen according to the stationary distribution $\mu=(\eta_v^T \nu_v)_{v\in V}$. Then for every $i\in\N$ and $k\in\N$, 
	\[\Pr\parenv{x_{[k]}=x_{i+[k]}}\leq \frac{|V| d^2}{\lambda^k},\]
	where $d=\max_{v,u\in V}\frac{\nu_v}{\nu_u}$ and $\lambda$ is the Perron eigenvalue of the adjacency matrix $A_G$.
\end{lemma}

\begin{IEEEproof}
	Recall that a path $\gamma$ is a sequence of edges $\gamma=(e_0,\dots, e_{k-1})$ such that for every $i\in [k-1]$, $e_i^t=e_{i+1}^b$. First we note that the probability of a specific path over the graph depends only on the start vertex, end vertex, and the length of the path. Indeed,  
	\begin{align}
	\label{eq:h11}
	\nonumber
	\Pr( (e_0,\dots,e_{k-1}))&= \mu_{e_0^b}\frac{P_{e_0^b,e_0^t}}{(A_G)_{e_0^b,e_0^t}}\frac{P_{e_1^b,e_1^t}}{(A_G)_{e_1^b,e_1^t}}\dots \frac{P_{e_{k-1}^b,e_{k-1}^t}}{(A_G)_{e_{k-1}^b,e_{k-1}^t}}\\ \nonumber
	&= \mu_{e_0^b}\frac{\nu_{e_0^b}}{\lambda \nu_{e_0^t}}\frac{ \nu_{e_1^b}}{\lambda \nu_{e_1^t}}\dots \frac{ \nu_{e_{k-1}^b}}{\lambda \nu_{e_{k-1}^t}}\\  &\stackrel{(a)}{=} \mu_{e_0^b}\frac{\nu_{e_0^b}}{ \nu_{e_{k-1}^t}}\frac{1}{\lambda^k},
	\end{align}
	where $(a)$ follows since $e_i^t=e_{i+1}^b$. Since the system is irreducible, $\mu,\nu,\eta$ are all positive. If we denote by $d$ the value $d=\max_{v,u\in V}\frac{\nu_v}{\nu_u}\geq 1$ 
	%and by $d'=\min_{v,u\in V} \frac{\nu_v}{\nu_u}>0$ then %
	we obtain 
	\[\Pr( (e_0,\dots,e_{k-1}))\leq \mu_{e_0^b}\frac{d}{\lambda^k}.\]
	For a sequence of edges $\gamma=(e_0,\dots, e_{k-1})$ we denote $L(\gamma)\eqdef L(e_0)L(e_1)\cdots L(e_{k-1})$ and denote by $\gamma_0$ the vertex $e_0^b$. We denote by $\Gamma$ the set of all paths and for $i\in \N$ we denote by $\Gamma^i$ the set of all paths of length $i$. Note that for a specific $w\in\Sigma^k$, 
	\begin{align*}
	\Pr(x_{[k]}=w)&=\sum_{\gamma\in\Gamma} \mathbb{1}_w(L(\gamma))\Pr(\gamma)\\
	&\leq \sum_{\gamma\in\Gamma} \mathbb{1}_w(L(\gamma))\mu_{\gamma_0}\frac{d}{\lambda^k}.
	\end{align*}
	
	Since the graph is deterministic, if $\gamma=(e_0,\dots,e_{k-1})$ is a path with $L(\gamma)=w$ then it is the only path with this labeling which starts at the vertex $e_0^b$. Thus, 
	\begin{align}
	\label{eq:helpp1}
	\Pr(x_{[k]}=w)\leq \sum_{v\in V}\mu_v\frac{d}{\lambda^k}\leq \frac{d}{\lambda^k}.
	\end{align}
	Since $\mu$ is the stationary probability vector, it is shift invariant, i.e., for $w\in\Sigma^k$ and $i\in\N$
	\[ \Pr(x_{i+[k]}=w)=\Pr(x_{[k]}=w).\] 
	Assume $i\in \N$ and write
	\begin{align}
	\label{eq:hh2}
	\nonumber
	\Pr\parenv{x_{[k]}=x_{i+[k]}}&=\sum_{w\in\Sigma^{i+k}}\mathbb{1}_{x_{[k]}}(x_{i+[k]})\Pr(x_{[k+i]}=w) \\ \nonumber
	&\leq \frac{d}{\lambda^{i+k}}\sum_{w\in\Sigma^{i+k}}\mathbb{1}_{x_{[k]}}(x_{i+[k]}) \\
	&\leq \frac{d}{\lambda^{i+k}}\cdot |\cB_i(S)|.
	\end{align}
	We now need to estimate the value $|\cB_i(S)|$. Note that $|\cB_i(S)|\leq |\Gamma^i|$. Since $\eta^T,\nu$ are left and right eigenvectors of $A_G$, respectively, for $i\in \N$ we may write
	\[\sum_{u\in V}\sum_{v\in V}(A_G^{i})_{u,v}\nu_v=\1 \cdot A_G^i \cdot \nu=\lambda^i \|\nu\|_1.\]
	Since $\|\nu\|_1\leq |V|\max_{v\in V} \mathset{\nu_v}$ and since $\forall v\in V,\; \nu_v\geq \min_{v\in V} \mathset{\nu_v}$ we obtain
	\[\1^T \cdot A_G^i \cdot \1=\sum_{u\in V}\sum_{v\in V}(A_G^{i})_{u,v}\leq |V|d\lambda^i.\]
	Since $|\Gamma^i|=\1 A_G^i\1$, plugging it in \eqref{eq:hh2} concludes the proof.
\end{IEEEproof}

We now state and prove the main result of this section.
\begin{theorem}
	\label{th:capmarkov1}
	Let $S$ be an irreducible constrained system presented by the graph $G$ with Perron eigenvalue $\lambda$. For every $n\in\N$ let $k=\lfloor a\log_{\lambda} (n)\rfloor$ with $a=(2+\epsilon)\log_{\lambda} 2$ with $\epsilon>0$. Then
	\[\ccap(\cX_{S,k})=\ccap(S).\]
\end{theorem}

\begin{IEEEproof}
    Let $\mathbb{P}(\cdot)$ denote the uniform probability over the length-$n$ sequences and note that $\frac{|\cX_{S,k}(n)|}{2^n}=\mathbb{P}(\cX_{S,k}(n))$. 
	Thus, $\ccap(\cX_{S,k})=1+\limsupup{n}\frac{1}{n}\log_2 \mathbb{P}(\cX_{S,k}(n))$.
	
	First note that $\cX_{S,k}(n)\subseteq \cB_n(S)$ which means that $\ccap(\cX_{S,k}(n))\leq \ccap(S)$. So we only need to show that $\ccap(\cX_{S,k}(n))\geq \ccap(S)$. 
	We show this using the first moment method. Assume that $S$ is presented by a graph $G=(V,E,L)$ with Perron eigenvalue $\lambda$ and an entropy maximizing transition probability $P$ with left Perron eigenvector $\eta^T$ and right Perron eigenvector $\nu$ normalized such that $\eta^T\cdot \nu=1$. Every sequence obtained according to $P$ belongs to $\cB(S)$. Denote by $\mu$ the stationary distribution of $P$. Let $w\in\Sigma^{n+k}$ obtained according to the Markov chain $P$ with initial symbol chosen according to $\mu$. 
	
	Let $\cI=\mathset{u=(u_0,u_1)\in [n]^2 ~:~ u_0\neq u_1}$ and for $u=(u_0,u_1)\in\cI$ we define $I_u=\mathbb{1}_{w_{u_0+[k]}}(w_{u_1+[k]})$ the indicator function for the event that the $k$-tuples that start in locations $u_0$ and $u_1$ are identical. 
	As shown in Lemma \ref{lem:shiftinv}, $\E\sparenv{I_u}\leq \frac{|V|d^2}{\lambda^k}$ where $d$ is given by the eigenvectors of $G$. 
	Applying Markov's inequality we obtain 
	\begin{align*}
	    \Pr\parenv{\sum_{u\in \cI} I_u \geq 1}&\leq \sum_{u\in \cI}\E\sparenv{I_u} \\ 
	    &\leq \sum_{u\in \cI}\frac{|V|d^2}{\lambda^k} \\
	    &\leq \frac{|V|d^2 n^2}{\lambda^k}.
	\end{align*}
	Since $k=a\log_2 (n)$ with $a=(2+\epsilon)\log_{\lambda} 2$ we obtain 
	\[\Pr\parenv{\sum_{u\in \cI} I_u \geq 1}\leq \frac{|V|d^2 n^2}{n^{2+\epsilon}}.\] 
	This, in turn, implies that 
	\[\Pr\parenv{\hat{\cW}_k(n+k)}=1-\Pr\parenv{\sum_{u\in \cI} I_u \geq 1}\geq 1-\frac{|V|d^2 n^2}{n^{2+\epsilon}}.\]
	Taking $n\to\infty$ we obtain 
	\begin{align}
	\label{eq:cap1}
	\limsupup{n}\frac{1}{n+k}\log_2 \Pr(\hat{\cW}_k(n+k))\geq 0.
	\end{align}
	Next, let $\mathbb{P}$ denote the uniform distribution over $\Sigma^{n+k}$. Note that in order to use the probability argument in order to estimate $|\hat{\cW}_k(n+k)|$, we need to use the uniform distribution (indeed, $|\cX_{S,k}(n+k)|=|\Sigma|^{n+k}\cdot \mathbb{P}(\cX_{S,k}(n+k))$ since $\mathbb{P}$ is the uniform distribution). We have 
	\[\frac{1}{n+k}\log_2 \abs{\cX_{S,k}(n+k)}= 1+\frac{1}{n+k}\log_2 \mathbb{P} \parenv{\cX_{S,k}(n+k)}.\]
	Now note that the probability denoted by $\Pr(\cdot)$ is not the uniform distribution but a distribution obtained by the Markov chain $P$ with stationary initial distribution $\mu$. To finish the proof we need to show that for $n$ large enough, $\Pr(\cdot)$ is almost uniform on the set $\cB_{n+k}(S)$.
	By the definition of $\cX_{S,k}(n+k)$ we have that $\mathbb{P} \parenv{\cX_{S,k}(n+k)}=\mathbb{P}\parenv{\cB_{n+k}(S)} \mathbb{P} \parenv{\hat{\cW}_k(n+k)\given \cB_{n+k}(S)}$. Hence, we obtain 
	\begin{align*}
	\frac{1}{n+k}\log_2 \abs{\cX_{S,k}(n+k)}&= 1+\frac{1}{n+k}\log_2 \mathbb{P}\parenv{\cB_{n+k}(S)}  \\
	&+ \frac{1}{n+k}\log_2 \mathbb{P} \parenv{\hat{\cW}_k(n+k)\given \cB_{n+k}(S)}.
	\end{align*}
	Note that 
	\[\ccap(S)=\log_2 \lambda=1+\limsupup{n}\frac{1}{n+k}\log_2 \mathbb{P}\parenv{\cB_{n+k}(S)}.\]
	Therefore, we have
	\begin{align}
	\label{eq:cap2}
	&\limsupup{n}\frac{1}{n+k}\log_2 \abs{\cX_{S,k}(n+k)}=\\ \nonumber 
	&\log_2 \lambda +\limsupup{n}\frac{1}{n+k}\log_2 \mathbb{P} \parenv{\hat{\cW}_k(n+k)\given \cB_{n+k}(S)}.
	\end{align}
	We claim now that 
	\begin{align}
	\label{eq:cap3}
	&\limsupup{n}\frac{1}{n+k}\log_2 \mathbb{P} \parenv{\hat{\cW}_k(n+k)\given \cB_{n+k}(S)} = \\ \nonumber 
	&\limsupup{n}\frac{1}{n+k}\log_2 \Pr \parenv{\hat{\cW}_k(n+k)}.
	\end{align}
	Showing this will finish the proof since plugging \eqref{eq:cap3} to (\ref{eq:cap2}), together with (\ref{eq:cap1}) yields
	\[\ccap(\cX_{S,k})\geq \log_2 \lambda= \ccap(\cB(S)).\]
	Note that (\ref{eq:cap3}) follows directly from Lemma \ref{lem:shiftinv}. Indeed,
	\begin{align*}
	\Pr \parenv{\hat{\cW}_k(n)}&=\sum_{w\in\hat{\cW}_k(n)} \Pr(\mathset{w}) \\ 
	&\leq \abs{\hat{\cW}_k(n+k)\cap \cB_{n+k}(S)}\frac{|V|d^2}{\lambda^{n+k}}.
	\end{align*}
	On the other hand, denoting $d'=\min_{u,v\in V}\frac{\nu_v}{\nu_u}$ we obtain from \eqref{eq:h11} that 
	\[\Pr(\hat{\cW}_k(n+k))\geq |\hat{\cW}_k(n+k)\cap\cB_{n+k}(S)| \min_{v\in V}\mu_{v}\frac{d'}{\lambda^{n+k}}.\]
	Thus, 
	\begin{align}
	\label{eq:cap4}
	&\limsupup{n} \frac{1}{n+k}\log_2 \Pr \parenv{\hat{\cW}_k(n+k)}=\\ \nonumber 
	&\limsupup{n}\frac{1}{n+k}\log_2 \abs{\hat{\cW}_k(n+k)\cap\cB_{n+k}(S)} -\log_2 \lambda.
	\end{align}
	Since $\mathbb{P}$ is the uniform probability, we have that 
	\[\mathbb{P}\parenv{\hat{\cW}_k(n+k)\given \cB_{n+k}(S)}=\frac{\abs{\hat{\cW}_k(n+k)\cap\cB_{n+k}(S)}}{|\cB_{n+k}(S)|}\] 
	which means that 
	\begin{align}
	\label{eq:cap5}
	&\limsupup{n}\frac{1}{n+k}\log_2 \mathbb{P}\parenv{\hat{\cW}_k(n+k)\given \cB_{n+k}(S)}= \\ \nonumber 
	&\limsupup{n}\frac{1}{n+k}\log_2 \abs{\hat{\cW}_k(n+k)\cap\cB_{n+k}(S)} -\log_2\lambda.
	\end{align}
	Combining \eqref{eq:cap4} with \eqref{eq:cap5} we obtain the wanted equality which concludes the proof.	
\end{IEEEproof}

\begin{remark}
Theorem \ref{th:capmarkov1} can be proved also using Lov\'asz local lemma but will yield a similar result, i.e., a similar value for $a$.
\end{remark}

\begin{example}
	In this example we consider the (inverted) $(0,1)$-RLL constrained $k$-repeat free sequences (the constrained system is denoted by $S$). Hence, we are interested in sequences for which every $k$-tuple appears at most once and also there are no consecutive ones. We define accordingly the set 
	\[\cX_{S,k}(n)=\mathset{w\in\Sigma^n ~:~ \fr^2_w(11)=0 \text{ and } w\in\hat{\cW}_k(n)}.\]
	%Let $F_i$ denote the $i$th Fibonacci number defined by the recursion $F_n=F_{n-1}+F_{n-2}$ with $F_0=F_{-1}=1$. It is well known that the number of length-$n$ sequences which obey the $(0,1)$-RLL constraint is given by $F_{n+1}+2F_n+F_{n-1}$ (see for example \cite[Example 4.4]{LinMar85}).
	
	We start by considering the adjacency matrix of the $(0,1)$-RLL system which is given by 
	\[A_G=\begin{bmatrix}
	1 & 1 \\
	1 & 0
	\end{bmatrix}.\]
	The Perron eigenvalue is $\lambda=\frac{1+\sqrt{5}}{2}$ and the corresponding eigenvectors are $\eta= \frac{1}{\sqrt{\lambda+2}}\begin{bmatrix} \lambda & 1 \end{bmatrix}, \nu=\frac{1}{\sqrt{\lambda+2}}\begin{bmatrix} \lambda\\ 1 \end{bmatrix}$ (note that $\lambda^2=\lambda+1$). The transition probabilities that maximize the entropy are given by 
	\[P=\begin{bmatrix}
	\frac{1}{\lambda} & \frac{1}{\lambda^2} \\
	1 & 0
	\end{bmatrix},\]
	with stationary distribution $\mu=\begin{bmatrix} \frac{\lambda+1}{\lambda+2} & \frac{1}{\lambda+2}\end{bmatrix}$. 
	By Lemma \ref{lem:shiftinv} we obtain that for every word $w\in S$ of length $n$, $\Pr(w)\leq \frac{2\lambda^2}{\lambda^n}$. By Theorem \ref{th:capmarkov1} the capacity $\ccap(\cX_{S,k}(n))=\log_2 (\lambda)$ when $k=\lfloor a\log_{\lambda} (n)\rfloor$ with $a=(2+\epsilon)\log_{\lambda} 2$.
\end{example}

\section{Multidimensional $k$-Repeat Free Patterns}
\label{sec:mdcap}
In this section we generalize the capacity results of Section \ref{sec:pre} to the multidimensional case. First, we generalize the relevant notations. Let $\N^d$ be the $d$-dimensional grid. For a vector $v=(v_0,\dots,v_{d-1})\in\N^d$, we define $[v]=[v_0]\times[v_1]\times\dots\times[v_{d-1}]$. We also use $\be_i$ the unit vector of direction $i$. For $n\in\N$, we denote by $[n]^d$ the $d$-dimensional cube of length $n$, i.e., $[n]^d=\otimes_{i=0}^{d-1}[n]$. Let $w\in\Sigma^{[n]^d}$ and let $v\in [n]^d$, we denote by $w_{v}$ the symbol located in the $v$ location. We also denote by $\Sigma^{*d}=\bigcup_{n\in\N}\Sigma^{[n]^d}$ the set of all $d$-dimensional finite cubes.

We now define the $d$-dimensional capacity and the empirical frequency.
\begin{definition}
	Let $\cL\subseteq\Sigma^{*d}$ be a system. The capacity of $\cL$ is denoted by $\ccap(\cL)$ and is defined as
	\[ \ccap(\cL)\eqdef \limsupup{n}\frac{1}{n^d}\log_{|\Sigma|} |\cL\cap\Sigma^{[n]^d}|.\]
\end{definition}
For a pattern $w\in\Sigma^{[n]^d}$ and for a set of coordinates $A\subseteq [n]^d$, we denote by $w_A$ the restriction of $w$ to the set $A$. We also denote by $|w|$ the side-length $n$ of $w$.

\begin{definition}
	Let $w\in\Sigma^{[n]^d}$ and $k\leq n$. The empirical frequency of $k$-patterns in $w$ is denoted by $\fr^k_w$ and is defined as follows. For a $k$-pattern $u\in\Sigma^{[k]^d}$, 
	\[\fr_w^k(u)\eqdef \frac{1}{(n-k+1)^d}\sum_{v\in [n-k+1]^d} \mathbb{1}_u \parenv{w_{v+[k]^d}}. \]
\end{definition}
For the measure $\fr^k_w$, the support of $\fr^k_w$, $\Supp (\fr^k_w)$, is the set of all $k$-patterns that appear in $w$.

\begin{example}
	Let $\Sigma$ be the binary alphabet and let 
	\begin{align*}
	w_1= \begin{bmatrix}
	0&1&1&0 \\
	1&0&0&0 \\
	1&0&1&0 \\
	1&1&1&1
	\end{bmatrix}, 
	w_2= \begin{bmatrix}
	1&1&0&0 \\
	1&0&1&0 \\
	1&0&0&1 \\
	1&1&1&1
	\end{bmatrix}.
	\end{align*}
	Let $k=2$ and let 
	\[u=\begin{bmatrix}
	1&0 \\
	0&1
	\end{bmatrix}.\]
	Note that $\fr^2_{w_1}$ is the empirical frequency of $2\times 2$ matrices in $w_1$. We have that 
	\begin{align*}
	\Supp{\fr^k_{w_1}}&=\Bigg\{\begin{bmatrix}0&1 \\ 1&0  \end{bmatrix},\begin{bmatrix}1&1 \\ 0&0  \end{bmatrix},\begin{bmatrix}1&0 \\ 0&0  \end{bmatrix}, \begin{bmatrix}1&0 \\ 1&0  \end{bmatrix},\begin{bmatrix}0&0 \\ 0&1  \end{bmatrix}, \\
	& \qquad \begin{bmatrix}0&0 \\ 1&0  \end{bmatrix},\begin{bmatrix}1&0 \\ 1&1  \end{bmatrix},\begin{bmatrix}0&1 \\ 1&1  \end{bmatrix}\Bigg\}.
	\end{align*}
	Also, $\fr^k_{w_1}(u)=0$ and $\fr^k_{w_2}(u)=\frac{2}{9}$.
\end{example}

A $d$-dimensional De-Bruijn system over the alphabet $\Sigma$ with $|\Sigma|=q$ is denoted by $\cB^d_q$ and is defined as the set of all De-Bruijn patterns (over $\Sigma$) of span $[k]^d$ for all $k\in\N$. In a notational form, a $d$-dimensional De-Bruijn system is the set 
\begin{align*}
&\cB^d_{q}= \\ 
&\mathset{w\in \Sigma^{*d} : \exists k\in\N\; s.t.\; \forall u\in\Sigma^{[k]^d}, \fr^k_w(u)=\frac{1}{(|w|-k+1)^n}}.
\end{align*}
In a similar fashion, we define the $d$-dimensional $k$-repeat free patterns.
\begin{definition}
	A pattern $w\in\Sigma^{[n]^d}$ is said to be length-$n$ $k$-repeat free if every $[k]^d$-tuple appears at most once. The set of length $n$-length $k$-repeat free patterns is denoted by 
	\[\cW^d_k(n)\eqdef \mathset{w\in\Sigma^{[n]^d} ~:~ \forall u\in\Sigma^{[k]^d},\; \fr^k_w(u)\leq \frac{1}{(n-k+1)^d}}.\]
\end{definition}
Note that $\cW^1_k(|\Sigma|^k+k-1)$ is exactly the set of all De-Bruijn sequences of span $k$. Moreover, if $k<\parenv{d\log_{|\Sigma|} (n-k+1)}^{\rfrac{1}{d}}$ then it holds that $\cW^d_k(n)=\emptyset$. Therefore, we are interested in studying the size of the set $\cW^d_k(n)$ where $k>\parenv{d\log_{|\Sigma|} (n-k+1)}^{\rfrac{1}{d}}$. Consider the uniform distribution over all $d$-dimensional patterns of length $n$, then 
\[|\cW^d_k(n)|=|\Sigma|^{n^d}\cdot \Pr\left({\cW^d_k(n)}\right).\] 
For $a>1$, we define the $d$-dimensional $k$-repeat free system as $\cW^d_a=\bigcup_{n\in\N} \cW^d_k(n)$, where $k^d=\floorenv{ad\log_{q} (n)}$. The capacity, in this case, is given by 
\begin{align}
\label{eq:c2}
\ccap(\cW^d_a)=1+\limsupup{n}\frac{1}{n}\log_{|\Sigma|} \Pr(\cW^d_k(n)).
\end{align}

Our main result in this section is stated in the following theorem, which is a generalization of Theorem~\ref{th:cap} for the $d$-dimensional case.
\begin{theorem}
	\label{th:cap2}
	Let $\Sigma$ be a finite alphabet of size $q$ then for all $a>1$, $\ccap(\cW^d_a)=1$. 
\end{theorem}

\begin{IEEEproof}
    The proof follows a similar line as the proof of Theorem \ref{th:cap}. Let $n\in\N$ and $k^d=a\log_q (n^d)$ (we assume for simplicity that $a\log_q n$ is an integer).
	For a number $1<\ell\in [n]$ we denote by $F_{\ell}$ the $d$-dimensional cube of length $\ell$, 
	\[F_{\ell}\eqdef \mathset{v\in \N^d ~:~ \|v\|_{\infty}\leq \ell-1}. \]
	Let $w\in\Sigma^{[n+k]^d}$ be a random word in which each coordinate is chosen uniformly and independently over $\Sigma$. For $u=(u_0,u_1)\in [n]^d\times [n]^d$ we denote by $I_u=\mathbb{1}_{w_{u_0+F_k}}(w_{u_1+F_k})$ the indicator function of the event that the $d$-dimensional cubes of length $k$ that start in positions $u_0$ and $u_1$ are identical. Let $\cI:=\mathset{u=(u_0,u_1)\in [n]^d\times [n]^d ~:~ u_0\neq u_1}$ and notice that we are interested in a lower bound on 
	\[ \Pr\parenv{\sum_{u\in\cI} I_u =0}.\] 
	Again, it is clear that if $u=(u_0,u_1),v=(v_0,v_1)$ are such that $u_0+F_k, u_1+F_k$ do not overlap $v_0+F_k$ or $v_1+F_k$, then $I_u,I_v$ are independent. In addition, we have $\Pr(I_u=1)=\frac{1}{q^{k^d}}$ for every $u\in \cI$. 
	
	We will use Lemma \ref{lem:LLL} with $\mathset{A_i ~:~ i\in [m]}=\mathset{(I_u=1) ~:~ u\in\cI}$ where we draw an edge $(u,v)$ between $u$ and $v$ if at least one of $u_0+F_k, u_1+F_k$ overlaps with $v_0+F_k, v_1+F_k$. Thus, every $u\in \cI$ has at most $2^{d+1}k^d n^d=2(2kn)^d$ neighbours. 
	
	We set the real numbers to be $x_u=\frac{1}{2(2k n)^d}$ for every $u\in \cI$. For $n$ large enough the condition of the lemma holds since  
	\[\prod_{(u,v)\in E} (1-x_v) =\parenv{1-\frac{1}{2(2k n)^d}}^{2(2k n)^d} \to  e^{-1}\]
	and since $\frac{1}{2 (2k n)^d}$ decreases at a lower rate than $\Pr(I_u=1)=\frac{1}{q^{k^d}}=\frac{1}{n^{ad}}$ with $a=(1+\epsilon)>1$. 
	Applying Lemma \ref{lem:LLL} we obtain 
	\[\Pr\parenv{w\in \cW_a^d}\geq \prod_{u\in \cI}\parenv{1-\frac{1}{2(2k n)^d}}.\] 
	Since $\parenv{1-\frac{1}{2(2k n)^d}}\leq 1$ and since $|\cI|\leq n^{2d}$ we obtain 
	\[\Pr\parenv{w\in \cW_a^d}\geq \parenv{1-\frac{1}{2(2k n)^d}}^{n^{2d}}.\] 
	The result follows after taking logarithm and dividing by $n^d$ since $\parenv{1-\frac{1}{2(2k n)^d}}^{n^{2d}}\sim \exp\parenv{-\frac{n^d}{\log (n^d)}}$.
\end{IEEEproof}

\section{Conclusion} \label{sec:conc}
%%%%%%%%%%%%%%%%%%%%%%%%
In this paper we consider $k$-repeat free sequences over a general alphabet, which generalize the well-known De-Buijn sequences. We calculate the capacity of 
the sequences for $k$ which is a function of the sequence's length. We also study the capacity of $k$-repeat free sequences with local constraints, imposed by 
a given irreducible constrained system, and the capacity of $d$-dimensional $k$-repeat free patterns. For the binary case, we also provide an efficient encoding and decoding scheme that achieves the capacity. 

As a future work, it will be interesting to find an efficient encoding and decoding scheme for $k$-repeat free sequences with local constraints. We believe that it is possible to modify our coding technique and to adjust it to this case. It is also interesting to find an efficient coding algorithm for the $d$-dimensional $k$-repeat free patterns.

\section{Acknowledgment}
The authors would like to thank the associate editor and
the anonymous reviewers, whose comments helped improve
the presentation of the paper.

\bibliographystyle{IEEEtran}
\bibliography{allbib}

\begin{thebibliography}{10}
\providecommand{\url}[1]{#1}
\csname url@samestyle\endcsname
\providecommand{\newblock}{\relax}
\providecommand{\bibinfo}[2]{#2}
\providecommand{\BIBentrySTDinterwordspacing}{\spaceskip=0pt\relax}
\providecommand{\BIBentryALTinterwordstretchfactor}{4}
\providecommand{\BIBentryALTinterwordspacing}{\spaceskip=\fontdimen2\font plus
\BIBentryALTinterwordstretchfactor\fontdimen3\font minus
  \fontdimen4\font\relax}
\providecommand{\BIBforeignlanguage}[2]{{%
\expandafter\ifx\csname l@#1\endcsname\relax
\typeout{** WARNING: IEEEtran.bst: No hyphenation pattern has been}%
\typeout{** loaded for the language `#1'. Using the pattern for}%
\typeout{** the default language instead.}%
\else
\language=\csname l@#1\endcsname
\fi
#2}}
\providecommand{\BIBdecl}{\relax}
\BIBdecl

\bibitem{Fre1982}
H.~Fredricksen, ``A survey of full length nonlinear shift register cycle
  algorithms,'' \emph{SIAM review}, vol.~24, no.~2, pp. 195--221, 1982.

\bibitem{FarSchBru15}
F.~Farnoud, M.~Schwartz, and J.~Bruck, ``A stochastic model for genomic
  interspersed duplication,'' in \emph{IEEE Int. Symp. Inf. Theory
  (ISIT)}.\hskip 1em plus 0.5em minus 0.4em\relax IEEE, 2015, pp. 904--908.

\bibitem{FarSchBru16}
------, ``The capacity of string-duplication systems,'' \emph{IEEE Trans. Inf.
  Theory}, vol.~62, no.~2, pp. 811--824, 2016.

\bibitem{EliFarSchBru16}
O.~Elishco, F.~Farnoud, M.~Schwartz, and J.~Bruck, ``The capacity of some
  p{\'o}lya string models,'' in \emph{IEEE Int. Symp. Inf. Theory
  (ISIT)}.\hskip 1em plus 0.5em minus 0.4em\relax IEEE, 2016, pp. 270--274.

\bibitem{EliFarSchBru18}
\BIBentryALTinterwordspacing
------, ``The capacity of some p{\'{o}}lya string models,'' \emph{CoRR}, vol.
  abs/1808.06062, 2018. [Online]. Available:
  \url{http://arxiv.org/abs/1808.06062}
\BIBentrySTDinterwordspacing

\bibitem{LouFarSchBru18}
H.~Lou, F.~Farnoud, M.~Schwartz, and J.~Bruck, ``Evolution of $ k $-mer
  frequencies and entropy in duplication and substitution mutation systems,''
  \emph{arXiv preprint arXiv:1812.02250}, 2018.

\bibitem{CheChrKiaNgu17}
Y.~M. Chee, J.~Chrisnata, H.~M. Kiah, and T.~T. Nguyen, ``Deciding the
  confusability of words under tandem repeats,'' \emph{arXiv preprint
  arXiv:1707.03956}, 2017.

\bibitem{JaiFarSchBru17}
S.~Jain, F.~Farnoud, M.~Schwartz, and J.~Bruck, ``Duplication-correcting codes
  for data storage in the dna of living organisms,'' \emph{IEEE Trans. Inf.
  Theory}, vol.~63, no.~8, pp. 4996--5010, 2017.

\bibitem{DolAna10}
L.~Dolecek and V.~Anantharam, ``Repetition error correcting sets: Explicit
  constructions and prefixing methods,'' \emph{SIAM Journal on Discrete
  Mathematics}, vol.~23, no.~4, pp. 2120--2146, 2010.

\bibitem{LenWacYaa17}
A.~Lenz, A.~Wachter-Zeh, and E.~Yaakobi, ``Bounds on codes correcting tandem
  and palindromic duplications,'' \emph{arXiv preprint arXiv:1707.00052}, 2017.

\bibitem{Wac2018}
A.~Wachter-Zeh, ``List decoding of insertions and deletions,'' \emph{IEEE
  Trans. Inf. Theory}, vol.~64, no.~9, pp. 6297--6304, 2018.

\bibitem{MarSki95}
D.~Margaritis and S.~S. Skiena, ``Reconstructing strings from substrings in
  rounds,'' in \emph{36th Annual Symp. Foundations of Computer Science}.\hskip
  1em plus 0.5em minus 0.4em\relax IEEE, 1995, pp. 613--620.

\bibitem{KiaPueMil2016}
H.~M. Kiah, G.~J. Puleo, and O.~Milenkovic, ``Codes for dna sequence
  profiles,'' \emph{IEEE Trans. Inf. Theory}, vol.~62, no.~6, pp. 3125--3146,
  2016.

\bibitem{BenMeySchSmiSto91}
B.~Manvel, A.~Meyerowitz, A.~Schwenk, K.~Smith, and P.~Stockmeyer,
  ``Reconstruction of sequences,'' \emph{Discrete Mathematics}, vol.~94, no.~3,
  pp. 209--219, 1991.

\bibitem{Sco97}
A.~D. Scott, ``Reconstructing sequences,'' \emph{Discrete Mathematics}, vol.
  175, no. 1-3, pp. 231--238, 1997.

\bibitem{BatKanKhaMcG2004}
T.~Batu, S.~Kannan, S.~Khanna, and A.~McGregor, ``Reconstructing strings from
  random traces,'' in \emph{15th Annual ACM-SIAM Symp. Disc. Algorithms}.\hskip
  1em plus 0.5em minus 0.4em\relax Society for Industrial and Applied
  Mathematics, 2004, pp. 910--918.

\bibitem{DudSch2003}
M.~Dud{\i}k and L.~J. Schulman, ``Reconstruction from subsequences,''
  \emph{Journal of Combinatorial Theory, Series A}, vol. 103, no.~2, pp.
  337--348, 2003.

\bibitem{AchDasMilOrlPan10}
J.~Acharya, H.~Das, O.~Milenkovic, A.~Orlitsky, and S.~Pan, ``On reconstructing
  a string from its substring compositions,'' in \emph{IEEE Int. Symp. Inf.
  Theory (ISIT)}.\hskip 1em plus 0.5em minus 0.4em\relax IEEE, 2010, pp.
  1238--1242.

\bibitem{AchDasMilOrlPan15}
------, ``String reconstruction from substring compositions,'' \emph{SIAM
  Journal on Discrete Mathematics}, vol.~29, no.~3, pp. 1340--1371, 2015.

\bibitem{GabMil18}
R.~Gabrys and O.~Milenkovic, ``Unique reconstruction of coded sequences from
  multiset substring spectra,'' in \emph{IEEE Int. Symp. Inf. Theory (ISIT)},
  2018, pp. 2540--2544.

\bibitem{ChaChrEzeKia17}
Z.~Chang, J.~Chrisnata, M.~F. Ezerman, and H.~M. Kiah, ``Rates of dna sequence
  profiles for practical values of read lengths,'' \emph{IEEE Trans. Inf.
  Theory}, vol.~63, no.~11, pp. 7166--7177, 2017.

\bibitem{Lev01}
V.~I. Levenshtein, ``Efficient reconstruction of sequences from their
  subsequences or supersequences,'' \emph{J.~Combin.~Theory Ser.~A}, vol.~93,
  no.~2, pp. 310--332, Feb. 2001.

\bibitem{Ukk1992}
E.~Ukkonen, ``Approximate string-matching with q-grams and maximal matches,''
  \emph{Theoretical computer science}, vol.~92, no.~1, pp. 191--211, 1992.

\bibitem{DeB1946}
N.~G.~D. Bruijn, ``A combinatorial problem,'' \emph{Koninklijke Nederlandse
  Akademie v. Wetenschappen}, vol.~49, no.~49, pp. 758--764, 1946.

\bibitem{LevYaa17}
M.~Levy and E.~Yaakobi, ``Mutually uncorrelated codes for dna storage,'' in
  \emph{IEEE Int. Symp. Inf. Theory (ISIT)}, 2017, pp. 3115--3119.

\bibitem{LevYaa18}
------, ``Mutually uncorrelated codes for dna storage,'' \emph{IEEE Trans. Inf.
  Theor.}, vol.~65, no.~6, pp. 3671--3691, 2019.

\bibitem{HurIsa93}
G.~Hurlbert and G.~Isaak, ``On the de bruijn torus problem,'' \emph{Journal of
  Combinatorial Theory, Series A}, vol.~64, no.~1, pp. 50--62, 1993.

\bibitem{HurIsa95}
------, ``New constructions for de bruijn tori,'' \emph{Designs, Codes and
  Cryptography}, vol.~6, no.~1, pp. 47--56, 1995.

\bibitem{Ma84}
S.~Ma, ``A note on binary arrays with a certain window property (corresp.),''
  \emph{IEEE Trans. Inf. Theory}, vol.~30, no.~5, pp. 774--775, 1984.

\bibitem{ErdLov75}
P.~Erd{\H{o}}s and L.~Lov{\'{a}}sz, ``Problems and results on 3-chromatic
  hypergraphs and some related questions,'' in \emph{Infinite and finite sets},
  A.~H. {\em et al.}, Ed.\hskip 1em plus 0.5em minus 0.4em\relax North-Holland,
  Amsterdam, 1975, pp. 609--628.

\bibitem{AloSpe00}
N.~Alon and J.~Spencer, \emph{The Probabilistic Method (2nd Edition)}.\hskip
  1em plus 0.5em minus 0.4em\relax John Wiley \& Sons, Inc., 2000.

\bibitem{MarRotSie2001}
B.~H. Marcus, R.~M. Roth, and P.~H. Siegel, ``An introduction to coding for
  constrained systems,'' \emph{Lecture notes}, 2001.

\bibitem{FreKes86}
H.~Fredricksen and I.~J. Kessler, ``An algorithm for generating necklaces of
  beads in two colors,'' \emph{Discrete Math.}, vol.~61, pp. 181--188, 1986.

\bibitem{Mor04}
E.~Moreno, ``On the theorem of fredricksen and maiorana about de bruijn
  sequences,'' \emph{Advances in Applied Mathematics}, vol.~33, no.~2, pp.
  413--415, 2004.

\bibitem{RusSavWan1992}
F.~Ruskey, C.~D. Savage, and T.~M.~Y. Wang, ``Generating necklaces,'' \emph{J.
  Algorithms}, vol.~13, no.~3, pp. 414--430, 1992.

\bibitem{KupVar16}
O.~Kupferman and G.~Vardi, ``Eulerian paths with regular constraints,'' in
  \emph{mfcs16}, ser. Leibniz International Proceedings in Informatics
  (LIPIcs), vol.~58, 2016, pp. 1--62.

\end{thebibliography}
\end{document}